\documentclass[11pt,twocolumn]{article}
\usepackage{natbib,latexsym,amsmath,epsfig,vmargin,rotating,float}
\usepackage{longtable,amssymb,amsthm,hhline}
\theoremstyle{plain}

\bibpunct{(}{)}{;}{a}{,}{,}
\setpapersize{USletter}
\setmarginsrb{.25in}{.25in}{.25in}{.25in}{0pt}{0mm}{0pt}{.25in}
\setmarginsrb{.75in}{.75in}{.75in}{.25in}{0pt}{0mm}{0pt}{.5in}

\begin{document}
\bibliographystyle{sysbio}


\twocolumn[
    \begin{@twocolumnfalse}    
\begin{center} 
\textbf{\large Properties of Consensus Methods for Inferring Species Trees from Gene Trees\\}
\textbf{\today}\\
\vspace{.25in}
\sc{James H. Degnan$^{1*}$ Michael DeGiorgio$^2$, David Bryant$^3$, and Noah A. Rosenberg$^{1,2}$}\\

\emph{\footnotesize $^1$Department of Human Genetics, 1241 E. Catherine Street, University of Michigan, Ann Arbor  48109-0618, USA\\}
\emph{\footnotesize $^*$Corresponding author, E-mail: jamdeg@gmail.com}\\
\emph{\footnotesize $^2$Center for Computational Medicine and Biology, 2017 Palmer Commons, 100 Washtenaw Avenue, University of Michigan, Ann Arbor, 48109-2218 USA}\\
\emph{\footnotesize $^3$Department of Mathematics, University of Auckland, Private Bag 29019, Auckland, New Zealand}\\
\vspace{.25in}
\end{center}

\begin{center}
\textbf{Abstract}
\end{center}

{\small Consensus methods provide a useful strategy for combining
information from a collection of gene trees.  An important application
of consensus methods is to combine gene trees to estimate a species tree.
To investigate the
theoretical properties of consensus trees that would be obtained from
large numbers of loci evolving according to a basic evolutionary
model, we construct consensus trees from independent gene trees that
occur in proportion to gene tree probabilities derived from coalescent
theory.  
We consider majority-rule, rooted triple ($R^*$), and greedy consensus trees
constructed from known gene trees, both in the asymptotic case as numbers
of gene trees approach infinity and for finite numbers of genes.
Our results show that for some
combinations of species tree branch lengths, increasing the number of
independent loci can make the majority-rule consensus tree more likely
to be at least partially unresolved and the greedy consensus tree less
likely to match the species tree. However, the probability that the
$R^*$ consensus tree has the species tree topology approaches 1 as the
number of gene trees approaches infinity.  Although the greedy
consensus algorithm can be the quickest to converge on the correct
species tree when increasing the number of gene trees, it can also be
positively misleading. The majority-rule consensus tree is not a
misleading estimator of the species tree topology, and the $R^*$
consensus tree is a statistically consistent estimator of the species
tree topology. Our results therefore
suggest a method for using multiple loci to infer the species tree topology,
even when it
is discordant with the most likely gene tree. \\}
\vspace{.25in}
\end{@twocolumnfalse}
]

The goal of many phylogenetic and phylogeographic studies is not the
estimation of the individual gene trees, but rather the estimation of
the species-level phylogeny or population history
\citep{felsenstein1988,takahata1989,maddison1997,nei2000}.  Among
methods that have been used to estimate species trees from data on
multiple loci, a popular approach has been to make use of sequences
concatenated across the loci.  In essence, this approach assumes that
all loci have the same gene tree, whose estimate is also used as the
estimated species tree. Because gene trees vary both locally and
across broad regions of organismal genomes
\citep{chen2001,pollard2006,hobolth2007}, sequence data from multiple
genes are expected to be the result of heterogeneous processes.
Multilocus data can be regarded as mixtures generated from different
branch lengths and mutation rates on gene trees as well as from
different gene tree topologies that may arise from
sources such as incomplete lineage sorting or hybridization.

As a result of these various sources of heterogeneity, 
concatenation can perform poorly when sequences are analyzed as
if they come from a single model.  Inferences may be inconsistent
\citep{kolaczkowski2004}, or the mixture generating the sequences
might not be identifiable \citep{matsen2007} even when sites are
generated from the same topology.  Similarly, when sites are generated
from different topologies but under the same mutation model, analyzing
the concatenated data can lead to misleading inferences
\citep{mossel2005,edwards2007,kubatko2007}.  It is therefore useful to
examine the behavior of other approaches in situations with a high
level of gene tree discordance.

One approach for estimating species trees that 
does not assume all loci reflect the same
underlying gene tree is consensus trees.  
However, relatively little is known about how
consensus algorithms are expected to perform when applied to trees
from multiple loci. 
We explore the properties of three consensus algorithms applied to
independent loci when gene tree discordance is the
result of incomplete lineage sorting.  In particular, we ask the
question: as the number of gene trees considered from different loci
increases, what is the probability that the consensus tree matches the
species tree topology?  

We focus on majority-rule, rooted triple ($R^*$), and greedy consensus
trees. A survey of these and other consensus methods can be found in
Bryant~(2003).  Majority-rule consensus trees consist of those clades
that occur more than 50\% of the time in a collection of trees.  (For
simplicity, we always use 50\% as the cut-off when referring to
majority-rule consensus, although any greater proportion could be used
instead.)  The $R^*$ consensus tree is the most resolved tree that is
compatible with a set of three-taxon statements (rooted triples), each
of which is the rooted triple occurring most often (for a given set of
three taxa) in a collection of trees on the same set of taxa.  A tree
containing these rooted triples can be constructed using an algorithm
such as the method in \citep{bryant2001}.  We use the convention that
if the set of rooted
triples is incompatible or if there is a tie for the most frequently
occurring rooted triple, the $R^*$ tree is declared unresolved or partially
unresolved for those taxa causing the incompatibility.  Greedy
consensus trees are constructed by sequentially adding one clade at a
time, the most frequently occurring clade that is compatible with
clades already included in the greedy consensus tree (breaking ties
randomly).  Greedy consensus trees are also sometimes called
``Majority rule extended'' \citep{fels:phylip}, or simply
``Majority-rule'' \citep{baum2007}, and the greedy consensus algorithm
is implemented in PHYLIP \citep{fels:phylip} and PAUP*
\citep{swofford:paup}.  For a given set of input trees, the greedy and
$R^*$ consensus trees are always refinements of the majority-rule tree
\citep{bryant2003}, but can refine the majority-rule tree in different
ways.

The three consensus methods considered in this paper exhibit different
behaviors when the number of genes increases.  We find that in
evolutionary models that generate sufficient gene tree discordance,
adding genes can increase the probability that the majority-rule
consensus tree is unresolved. However, this unresolved tree is
compatible with the species tree in the sense that one of its
refinements has the species tree topology.  We call sets of branch
lengths leading to this lack of resolution \emph{unresolved zones}.
Also, as the number of independent, known gene trees increases, the
$R^*$ tree becomes fully resolved and matches the species tree.
However, greedy consensus trees, which are always resolved, can be
misleading in the sense that adding more genes can be more likely to
result in a tree that does not match the species tree.  We use the
term \emph{too-greedy zone} to denote the set of 
species tree branch lengths
for which greedy consensus trees constructed from infinitely many loci
disagree with the species tree.  This is analogous to the
\emph{anomaly zone} \citep{degnan2006}, the set of branch
lengths for which the most probable gene tree does not match the
species tree.  In the case of four-taxon asymmetric species trees, the
too-greedy zone is a subset of the anomaly zone.


In this paper, we first show some four-taxon examples of 
consensus trees when the number of loci approaches infinity but branch
lengths in the species tree vary. This is followed by derivations for
four-taxon trees
of the unresolved zones for majority-rule consensus trees and the too-greedy
zone for greedy consensus trees. The  
main results of the paper (Theorems 1, 3, 4, and 5)
give different results for the limiting behavior of the
three consensus methods used. 
Finally, we consider the same consensus methods with
finitely many loci sampled, including some examples with three and
four taxa.

\begin{center}
{\sc The Multispecies Coalescent}
\end{center}

We use the term ``multispecies coalescent'' for the model in
which coalescent
processes occur in each branch of a species tree and for which all possible
coalescent events within a branch are equally likely. This is the
model that has previously been used to calculate probabilities of gene
trees in species trees
\citep{tajima1983,pamilo1988,takahata1989,rosenberg2002,degnan2005}.
This model assumes that the genes from the different species are
orthologous, that there is no recombination or horizontal gene
transfer within the genes of interest, and that natural selection is
not acting on these genes.  This model also assumes that population
sizes are constant within species tree branches (although not
necessarily across branches) and that populations are panmictic.

\begin{center}
{\sc Definitions}
\end{center}

Unless otherwise noted, we use ``gene tree'' to refer to a gene tree
topology, and ``species tree'' to refer to a species tree topology
with internal branch lengths specified.  Because
two or more lineages in a population are needed for a coalescence to
occur, lengths of external branches
(those leading to tips of the species tree) do not affect
probabilities of gene tree topologies when only one lineage is considered per
species.  Branch lengths on species trees are measured in \emph{coalescent
units}, the number of generations divided by the effective population
size (twice the effective population size for diploids \citep{hein2005}).

Nodes on gene trees correspond to coalescent events.  For example if a
node on a gene tree is the root of the subtree ((AB)C), this node
corresponds to the coalescent event that joins the lineage ancestral to (AB)
with the lineage ancestral to C, where (AB) itself
represents the coalesced lineage combining the lineages from taxa A and B.
We say that (AB) is a lineage ``containing'' A and B.  
We additionally say that two taxa ``join'' or ``are joined'' on a branch $b$ 
if the lineages (i.e.~clades) containing those taxa coalesce on branch
$b$.  For example, if (AB) and C coalesce on branch 3, then A and C
``join'' on branch 3.  Clades with only two taxa (on either species or
gene trees) are called \emph{cherries}.
We use the same letter (such as A, B, etc.) to refer to both a
taxon and to the gene lineage sampled from that taxon.

We use the notation (AB)C for the three-taxon statement (rooted
triple) that the most recent common ancestor (MRCA) of gene lineages A
and B on a species tree is not an ancestor of C.  This notation is
similar to the notation for a three-taxon tree but does not have the
outer set of parentheses.  If a given species tree (with topology and
internal branch lengths specified) is $\sigma$, then
$P_{\sigma}[\cdot]$ indicates probabilities of events for gene
lineages when $\sigma$ is the species tree.  For example,
$P_{\sigma}[\text{(AB)C}]$ and $P_{\sigma}[\text{((AB)C)}]$ are used to
indicate the probabilities of the rooted triple (AB)C and the gene tree
((AB)C), respectively.
The expression $P_{\sigma}[\text{\{ABC\}}]$ is used to denote
the probability that \{ABC\} is a clade on the gene tree.

Because we frequently refer to time looking backwards starting from the
present, we use ``before'' and ``first'' to mean ``more recently''
and ``most recently'', and we use ``more anciently than''
in the usual sense of looking at time from the past to the present.

\begin{center}
{\sc Asymptotic Consensus Trees}
\end{center}

Consensus trees are used to summarize a set of trees defined on the
same set of taxa.  A consensus
algorithm takes the trees as inputs, so that the method of producing the
input trees is not part of the consensus algorithm. Typically the
trees summarized might be estimated trees such as those that are obtained from
separate genes, different models, or different bootstrap samples.  In all of
these cases, the consensus tree is a function of some data set and is
therefore a statistic \citep{casella1990}.

Using gene tree probability distributions, we can also compute the
consensus tree that would be returned in the limit as the number of
gene trees approaches infinity. This calculation assumes that these
gene trees are correctly estimated, independent, and generated by the
multispecies coalescent model.  In this setting, the proportion of
occurrences for a gene tree topology asymptotically approaches its
probability under the multispecies coalescent model as the sample size
(the number of independent loci) approaches infinity.  


Consensus
trees obtained from these asymptotic proportions are not functions of
data, and are therefore not statistics. Instead they are properties
solely of gene tree probability distributions. These in turn are
functions of the species tree, which we can consider to be a parameter
for a gene tree distribution \citep{degnan2005}.  Intuitively, we can
also think of a consensus tree computed from gene tree probabilities
under the multispecies coalescent as the consensus tree that would be
obtained from an infinite number of independent, correctly inferred
gene trees.  

We define an \emph{asymptotic consensus tree} for a species
tree to be the tree topology that would be obtained if a consensus
algorithm had input gene trees in proportion to their probabilities
(under the multispecies coalescent model).  
We note that under the multispecies coalescent model
that we are considering, every gene tree topology has positive
probability given any species tree, and therefore every gene
tree is included in
the consensus algorithm.  Consequently, methods 
such as Adams and strict consensus \citep{bryant2003,felsenstein2004}---which preserve information shared by all input trees---result in star trees when
probabilities under the multispecies coalescent are used.  As 
more gene trees are sampled, the
probability approaches zero that there is strict agreement for the
relationships for any subset of taxa.  We therefore
focus on three consensus algorithms that do not require strict
agreement.

For each of these algorithms, we first characterize the asymptotic
consensus trees for three and four taxa.  We also prove general
theorems about these trees for arbitrary numbers of taxa.  We then
return to the three- and four-taxon cases and consider the approach to
the asymptotic consensus tree based on finitely many loci.

The majority-rule asymptotic consensus tree (MACT) can be determined
by listing the probability of monophyly for each subset of taxa.  If a
subset of taxa appears on the list with probability greater than
1/2, then that group is contained in the MACT.  This is the same
method traditionally used to determine majority-rule consensus trees,
but here we use theoretical probabilities rather than observed
proportions.

Similarly, the $R^*$ asymptotic consensus tree (RACT) can be
determined by calculating the probability of each of the three
possible rooted triples for each of the $\binom{n}{3}$ subsets of
three taxa.  The RACT then consists of those rooted triples that have
the highest probability for each subset of three taxa.  For any three
taxa and a strictly bifurcating species tree, the rooted triple
corresponding to the species tree is always the most probable (see
Proposition~2 below)---i.e., there are no ties.  The set of rooted
triples for all ${n \choose 3}$ subsets of three taxa uniquely
identifies the species tree Steel~(1992, Prop.~4)\nocite{steel1992}; thus the RACT is always uniquely
identified and fully resolved under the multispecies coalescent model.

\begin{figure}[!t]
\begin{center}
\includegraphics[width=.49\textwidth,height=.49\textwidth]{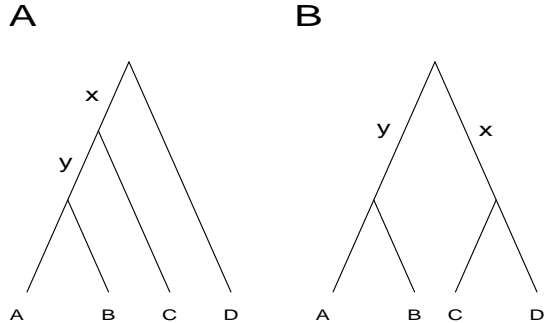}
\end{center}
\caption{Four-taxon species trees with internal branch lengths $x$ and $y$.}\label{F:plot1}
\end{figure}

The greedy asymptotic consensus tree (GACT) for $n$ taxa can be obtained
by ranking probabilities of the $2^n-n-1$ clades with two or more taxa.  
The most probable clade is
incorporated into the consensus tree, and then the list of clade probabilities
is updated by removing any clades incompatible with those already in
the tree.
This process is repeated until the tree is fully resolved, randomly
picking clades in the case of ties.

The three types of asymptotic consensus trees---MACT, RACT,
and GACT---are purely mathematical functions of gene tree
probabilities.  They are therefore properties of species trees.
Consensus trees constructed from finitely many loci under different
consensus algorithms are random variables, and are increasingly likely
to match their asymptotic counterparts as the number of loci
approaches infinity.

\phantom{aa}

\begin{center}
\emph{Examples}
\end{center}

Examples which illustrate the construction of asymptotic consensus trees for
the three methods in this paper are shown in Table~\ref{T:consensus},
which lists probabilities of each gene tree for four taxa, for several
sets of branch lengths on the species tree in Figure~1A.  Also
listed are probabilities for two- and three-taxon clades, and
probabilities for the 12 rooted triples.  For four taxa, there are six
possible cherries and four possible three-taxon
monophyletic groups.  Note that because the cherries are not
mutually exclusive, their probabilities sum to more than one.  Also,
because it is possible for a tree to not have any three-taxon
monophyletic groups, the sum of the probabilities for subsets of three
taxa is less than one.

For the examples in Table~1, majority-rule consensus returns each of
the four possible trees illustrated in Figure~2A.  Greedy consensus
returns the matching tree for all examples in the table, except when
$(x,y) = (0.05,0.05)$, in which case it returns ((AB)(CD)). This
topology is also the most probable gene tree for those branch lengths.
$R^*$ consensus is the only consensus method considered which returns
the matching tree for all branch lengths used.  From Theorem~3 this
result is not limited to the example chosen, but applies to any branch
lengths and any binary species tree.

As an example from the table, we see that if the species tree has
topology (((AB)C)D) and has $x = 0.6$ and $y = 0.4$, then the groups
\{AB\} and \{ABC\} both occur with probability greater than 1/2, and
\{CD\} occurs with probability less than 1/2.  Thus the MACT for this
species tree has the topology (((AB)C)D), since this is the only
four-taxon topology which has exactly the monophyletic groups \{AB\}
and \{ABC\}.  Both probabilities are only slightly larger than 1/2,
however, so in a small sample of correctly inferred trees, it is
likely that either \{AB\} or \{ABC\} would occur less than 50\% of
the time, or that \{CD\} would occur more than 50\% of the time. In
these cases, the majority-rule consensus tree would be unresolved or
would not match the species tree.  

For the greedy consensus algorithm, we would select the \{AB\} clade
to be in the tree (because it is the most probable other than
\{ABCD\}), and then eliminate all clades except \{CD\}, \{ABC\}, and
\{ABD\} from consideration since these other clades are incompatible
with \{AB\}.  From the three remaining clades, \{ABC\} is the most
probable---hence the GACT has clades \{AB\} and \{ABC\}, which means
that (((AB)C)D) is the GACT.  For the $R^*$ consensus algorithm, the
most probable rooted triples for each set of three taxa are: (AB)C,
(AB)D, (AC)D, and (BC)D.  Since (((AB)C)D) is the only tree for these
taxa that is compatible with these rooted triples, $R^*$ also returns the
matching tree.

Choosing the branch lengths to be $(x,y) = (0.4, 0.6)$ (Table~1, second branch length column), illustrates that
the behavior of MACTs is sensitive to the order of the branch lengths.
Switching the lengths for $x$ and $y$ can change whether the MACT is
fully resolved.  For this tree, most (about 62\%) gene trees are
expected to have an \{AB\} clade, so this clade is very likely to be in the
majority-rule consensus tree for a large enough number of gene trees;
however, less than 46\% of trees are expected to have \{ABC\} in a
monophyletic group, so the MACT does not have \{ABC\}
as a clade.  Since no other group is monophyletic with probability
greater than 1/2, this MACT is not fully resolved, and is ((AB)CD).
Note that this lack of resolution is a theoretical limitation of majority-rule
consensus and occurs even though the species tree and gene trees are fully
resolved (there are no ``hard'' polytomies). The lack of resolution is also not due to insufficient information---in other words, the lack
of resolution cannot be overcome by collecting more data (there are no ``soft'' polytomies).  

When the branch lengths are $(x,y) = (0.8, 0.3)$ (Table~1, third
branch length column), majority-rule consensus returns the other
partially resolved tree, ((ABC)D).  For the branch lengths $(x,y) = (0.3,0.3), (0.1,0.1), (0.05,0.05)$ (columns four through six), 
since no monophyletic subset of taxa has probability greater
than 1/2, the MACTs for this species tree are star phylogenies.    
When the branch lengths
are $(x,y) = (0.1,0.1)$ and $(x,y)=(0.05,0.05)$, ((AB)(CD)) 
is the most probable gene tree, although it does not match the 
species tree.  Gene trees that are more probable than the gene tree
matching the species tree are called \emph{anomalous gene trees} \citep{degnan2006}.  When $(x,y) = (0.3,0.3)$, no anomalous gene trees occur, so this example
illustrates that unresolved majority-rule consensus trees can arise
even when there are no anomalous gene trees.
When $(x,y) = (0.05,0.05)$, the most probable clade is \{AB\}, which has
probability 0.275, so it is included in the greedy consensus tree.
The second most probable clade compatible with \{AB\}, however, is
\{CD\}, which has probability 0.212, and thus the greedy consensus
tree is ((AB)(CD)), which does not match the species tree.

We now describe asymptotic consensus trees for more general sets of branch
lengths, considering three- and four-taxon trees as well as trees with
arbitrary numbers of taxa.

\begin{center}
\emph{Majority-rule Consensus}
\end{center}

\begin{figure*}[!t]
\begin{center}
\includegraphics[width=.49\textwidth,height=.49\textwidth]{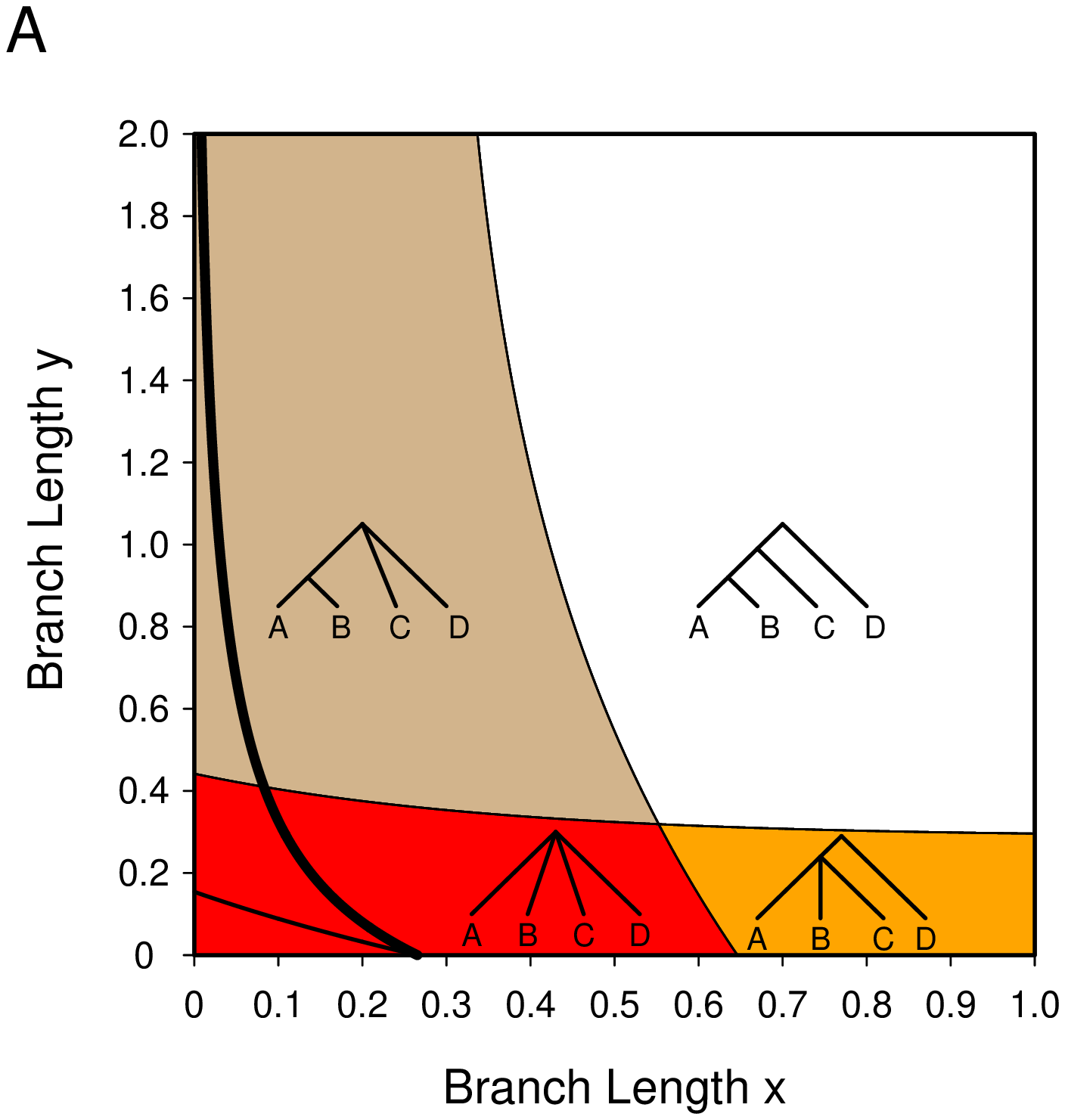}
\includegraphics[width=.49\textwidth,height=.49\textwidth]{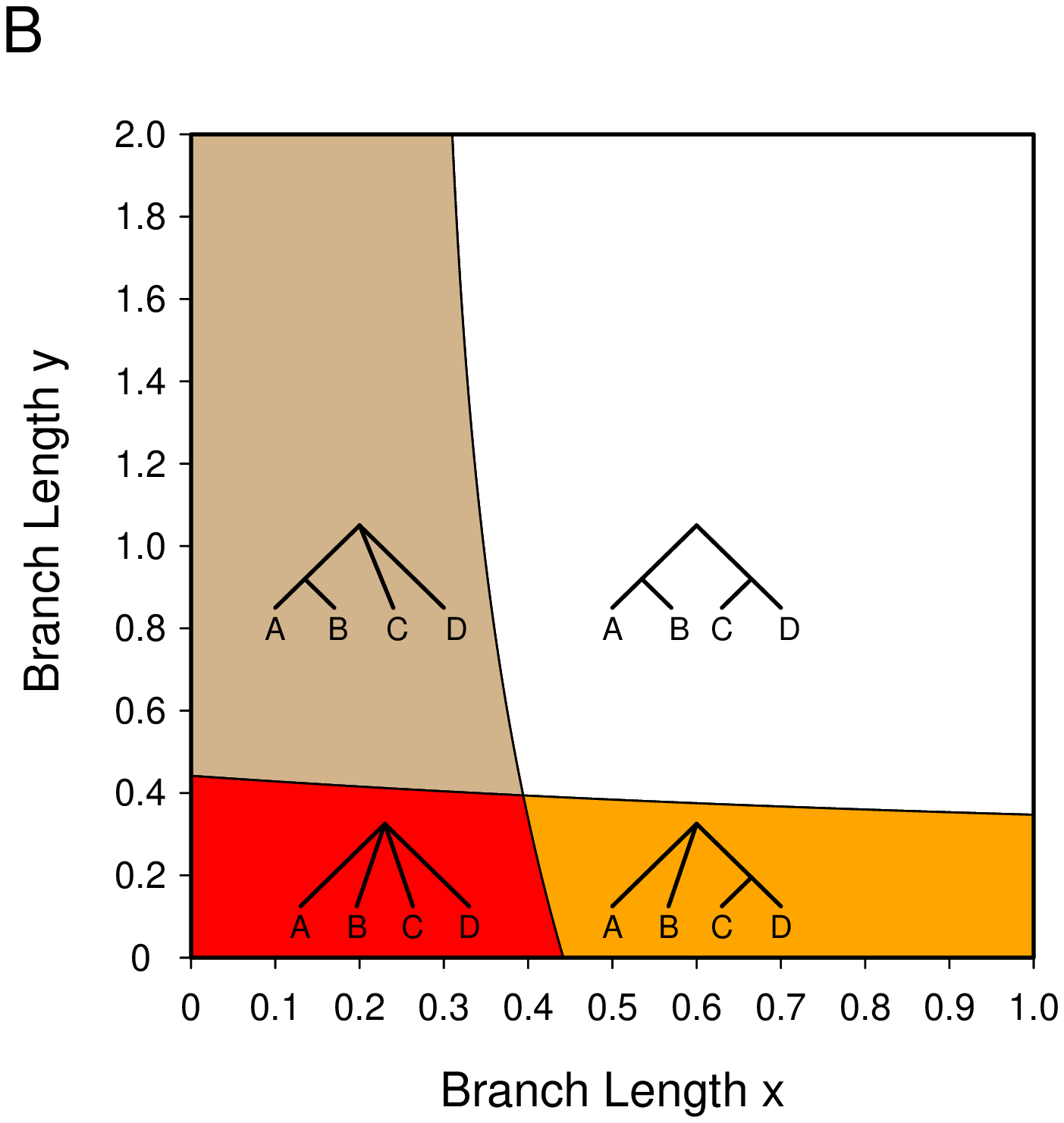}
\end{center}
\raggedright
\caption{Unresolved zones.  The shaded regions are different
areas of the unresolved zones leading to different unresolved 
majority-rule consensus trees.
(A) The species tree is (((AB)C)D). A star tree is the limiting consensus tree for the red
region, and the orange region corresponds to the tree with the \{ABC\} clade unresolved.  For comparison, the anomaly zone is also
plotted as the area under the heavy, dark curve.  The anomaly zone
cuts across two regions of the unresolved zone, and the area under the line
starting from $(x,y) = (0,0.154)$ which
creates the approximately triangular region is
the part of the anomaly zone with three anomalous gene trees. (B) The species tree
is ((AB)(CD)). The unresolved zone in this case is similar in size to that of (a), but there is no anomaly zone for this species tree. }\label{F:LOC}
\end{figure*}

Three taxa.---For the case of three-taxon trees, the MACT is
resolved if the probability of the matching tree is greater than
1/2.  Using the well-known probability of congruence for a gene tree
given a three-taxon species tree,
$1-(2/3)e^{-T}$ \citep{nei1987}, where
$T$ is the length of the one internal branch, this probability
is greater than 1/2 if $T > \log(4/3) \approx 0.28768$.  If the
internal branch length is less than this value, then increasing the number
of independent gene trees also increases the probability that the trees
do not produce a resolved majority-rule consensus tree, even though the
matching gene tree is more likely than any other gene tree.

Four taxa.---For four-taxon trees, the branch lengths needed for a
clade to be in the MACT can be obtained by setting the probability of
the clade to be greater than 1/2 and solving for branch length $y$ in terms
of branch length $x$.  These
clade probabilities are functions of gene tree probabilities and are
listed in Table~1.  The model four-taxon trees are shown in
Figure~\ref{F:plot1}.

Details for deriving conditions for clades to be in the MACT are given
in Appendices~1 and 2.  First we consider the species tree with topology
(((AB)C)D).  Following Figure~1, let $x$ be the length of the branch (in coalescent units)
ancestral to A and B, but not C, and let $y$ be the length of the other internal branch.  
Then \{ABC\} is a clade in the MACT if and only if
\begin{equation}\label{E:1}
x  > \log (4/3) \hspace{.5cm} \text{and} \hspace{.5cm} y > \log\left[\frac{2e^{2x}-1}{3e^{3x}-4e^{2x}}\right],
\end{equation}
and \{AB\} is a clade if and only if
\begin{equation}\label{E:2}
y > \log\left[\frac{12e^{3x} + 2}{9e^{3x}}\right].
\end{equation}

These two conditions partition the space of branch lengths into the four
possible MACTs for this species tree (Fig.~\ref{F:LOC}A), where
$x = \log (4/3) \approx 0.28768$ is a vertical asymptote.  The
MACT is:\\

\begin{align*}
&\text{(((AB)C)D)}\; \text{ if (\ref{E:1}) and (\ref{E:2}) both hold,}\\
&\text{((ABC)D)}\phantom{\text{()}}\;\text{ if (\ref{E:1}) holds and (\ref{E:2}) fails,}\\
&\text{((AB)CD)}\phantom{\text{()}}\;\text{ if (\ref{E:1}) fails and (\ref{E:2}) holds,}\\
&\text{(ABCD)}\phantom{\text{(())}}\;\text{ if (\ref{E:1}) and (\ref{E:2}) both fail.}\\
\end{align*}

Similarly, if the species tree is ((AB)(CD)), with $y$ denoting the length of the
branch ancestral to (AB) and $x$ denoting the length of the other internal branch,
then (AB) is a clade in the MACT if and only if
\begin{equation}\label{E:3}
y > \log\left[\frac{12e^x+2}{9e^x}\right],
\end{equation}
and (CD) is a clade in the MACT if and only if
\begin{equation}\label{E:4}
x > \log (4/3) \hspace{.5cm} \text{and} \hspace{.5cm} y > \log\left[\frac{2}{9e^{x}-12}\right].
\end{equation}
Again, these two conditions partition the branch length space into four regions,
one for each of the possible MACTs (Fig.~\ref{F:LOC}B), and $x = \log (4/3)$ is also a vertical asymptote for this graph. The MACT is:\\

\begin{align*}
&\text{((AB)(CD))}\; \text{ if (\ref{E:3}) and (\ref{E:4}) both hold,}\\
&\text{((AB)CD)}\phantom{\text{()}}\;\text{ if (\ref{E:3}) holds and (\ref{E:4}) fails,}\\
&\text{(AB(CD))}\phantom{\text{()}}\;\text{ if (\ref{E:3}) fails and (\ref{E:4}) holds,}\\
&\text{(ABCD)}\phantom{\text{(())}}\;\text{ if (\ref{E:3}) and (\ref{E:4}) both fail.}\\
\end{align*}

Arbitrarily many taxa.---Because equations (1)--(4) characterize all
possible MACTs for four taxa, it follows that four-taxon MACTs are
never misleading in the sense that a four-taxon MACT never has a clade
that is not a clade in the species tree.  Due to lack of resolution,
however, the MACT may fail to have clades that are in the species
tree.  Although we have obtained this result by explicit computation
for the four-taxon case, the result holds for larger trees:\\

\vspace{.25in}\textbf{Theorem 1}. (i) The majority-rule asymptotic
consensus tree does not have any clades not on the species tree. (ii)
For all species tree topologies with $n \ge 3$ taxa, there exist branch lengths
for which the majority-rule asymptotic consensus tree is not fully
resolved.
\vspace{.25in}

The proof of the first part of Theorem 1 is provided in the section on
$R^*$ trees below since it is a consequence of the consistency of
$R^*$ consensus (Theorem 3).  
The second part of Theorem 1 follows for the three-
and four-taxon cases from the calculations above.  For larger trees,
the second part of Theorem 1 follows from the inconsistency of greedy
consensus (Theorem 5) and the fact that greedy consensus trees are
refinements of majority-rule trees.

The plots in Figure~\ref{F:LOC} are analogous to the anomaly zone, the
region in branch length space in which the most likely gene tree does
not match the species tree \citep[Fig.~2]{degnan2006}.  Note that the region
of parameter space in which MACTs are not fully resolved (and
therefore do not fully recover the species tree) is considerably
larger than the anomaly zone.  For example, when we set $x=y$ for the
four-taxon asymmetric tree, the largest value of $x$ that is still in the
anomaly zone is approximately 0.1568 (Degnan and Rosenberg, 2006); but for
majority-rule consensus, $x=y=0.345$ is approximately the largest
value for which $x=y$ and the MACT is fully unresolved, and
$x=y=0.507$ is the largest value for which the MACT is partially
unresolved, equaling ((AB)CD).  For the symmetric four-taxon tree,
the values $x=y=0.394$ result in a star consensus tree. 
This is somewhat
surprising since these values result in the partially resolved
tree ((AB)CD) for the asymmetric species tree.
For the asymmetric four-taxon species tree, the anomaly zone
is a subset of the zone in which the MACT is unresolved. 
For the symmetric species
tree, the MACT is unresolved, 
but there is no anomaly zone.  For four
taxa, it is always the case that if a species tree has an anomalous
gene tree, it does not have a fully resolved MACT.

\begin{center}
$R^*$ \emph{Consensus}
\end{center}

Three taxa.---In the case of three taxa, we note that the greedy and
$R^*$ algorithms are equivalent when there are infinitely many loci.
For both algorithms, the most frequently occurring clade also
determines a three-taxon statement.  In the asymptotic case, there is
a uniquely occurring most frequent tree. This tree has probability
$1-(2/3)e^{-T} > 1/3$ (where $T$ is the one internal branch length),
and the other two trees each have probability $(1/3)e^{-T} < 1/3$.  Thus, for
the three-taxon case, as the number of loci approaches infinity, the
probability that the matching gene tree is the most frequent
approaches 1.

Arbitrarily many taxa.---We show that $R^*$ consensus trees are
consistent estimators of species tree topologies. This consistency is
based on the fact that for any set of three taxa, the rooted triple in the
species tree is the highest-probability rooted triple in the gene tree
distribution.

\vspace{.25in} \textbf{Lemma 2}.  Let $\sigma$ be the species
tree where $S$ is the set of taxa on $\sigma$.  For any A, B, C
$\in S$, if $\sigma$ has the grouping $\text{(AB)C}$, then
$P_{\sigma}[\text{(AB)C}] > P_{\sigma}[\text{(AC)B}]$.
\vspace{.25in}

\emph{Proof}. Let $\mathcal{J}$ be the set of branches on $\sigma$ on
which A and B can join (i.e., either the lineages A and B or the lineages
containing A and B can coalesce in $\mathcal{J}$), but on which A and
C cannot join.  Note that $\mathcal{J}$ is nonempty and that any
branch in $\mathcal{J}$ is an ancestor of A and B, and not an ancestor
of C.  Let $\mathcal{K}$ be the set of branches on which 
gene lineages A and C can
join.  Any branch in $\mathcal{K}$ is an ancestor of A and
C.  Since (AB)C is a rooted triple, any ancestor of A and C is also an
ancestor of B.  Thus for any branch $k \in \mathcal{K}$, if none of the
lineages A,
B, and C have joined, they are free to do so on $k$.  The
probability that A and B join on a branch in $\mathcal{J}$ is
positive. If A and B do not join in $\mathcal{J}$, then the
probabilities that A and B, A and C, and B and C are the first two of
A, B, and C to join in $\mathcal{K}$ are equal since all pairs of
lineages in a population are equally likely to coalesce.  Thus
$P_{\sigma}[\text{(AB)C}] > P_{\sigma}[\text{(AC)B}]$. $\square$

\vspace{.25in} \textbf{Theorem 3}. For a species tree $\sigma$, the
$R^*$ asymptotic consensus tree has the same topology as $\sigma$.
\vspace{.25cm}

\emph{Proof}.  By Lemma~2, any rooted triple in the species tree has
higher probability in the gene tree distribution than the other two
rooted triples for the same set of three taxa.  Thus, the set of
rooted triples from which the $R^*$ tree is constructed is exactly the
set of $\binom{n}{3}$ rooted triples in the species tree, where $n$ is
the number of taxa.  From Steel~(1992)\nocite{steel1992}, a
tree topology is uniquely specified by its set of rooted triples, from
which it follows that the only tree topology containing the
$\binom{n}{3}$ triples is the topology of the species tree
itself. $\square$

\vspace{.25in}
\emph{Proof of Theorem 1.} (i) This result follows from Proposition~3 and Theorem 2.14 of
Bryant~(2003), according to which every clade in the majority rule
consensus tree is in the $R^*$ tree.
Because the MACT and RACT are the majority-rule and $R^*$ consensus
trees applied to coalescent gene tree probabilities, 
every clade in the MACT must
appear in the RACT.  Because in the limit of
infinitely many gene trees, the $R^*$ tree is fully resolved, it
follows that if the MACT has one or more multifurcations, the $R^*$
tree is one of the possible resolutions of the MACT.  Because
the $R^*$ tree has the same topology as the species tree (Theorem 3), the MACT
either has the species tree topology or one its resolutions has the
same topology as the species tree. 
$\square$

Theorem 3 describes the RACT, which is a mathematical function of
gene tree probabilities, and therefore of species tree branch lengths.
When an $R^*$ consensus tree is computed from data, however, it has
some probability of not matching the species tree.  For an estimator
of a parameter to be statistically consistent, the probability that it
gets arbitrarily close to the parameter must approach 1 as the sample
size approaches infinity.  Theorem 4 describes the behavior of the
$R^*$ consensus tree constructed from data when the sample size
approaches infinity.

\vspace{.25in}
\textbf{Theorem 4}.  $R^*$ consensus is statistically consistent. 
\vspace{.25in}

The proof of Theorem~4 uses a generalized version of Bonferroni's
inequality, according to which if there are $k$ events each with
probability $p=1-q$, the probability that they all occur is greater
than or equal to $1-kq$ \citep[p. 63]{ross1998}.

\emph{Proof}. It must be shown that for any $\varepsilon > 0$, there
exists $k$ such that if there are at least $k$ independent gene trees,
the probability is greater than $1-\varepsilon$ that all rooted
triples in the species tree are also the most frequently occurring
rooted triples for each set of three taxa in the collection of gene
trees.  Let the species tree be $\sigma$ with taxon set $S$.  For $n$
taxa, there are $\binom{n}{3}$ sets of three taxa in $S$.  Let A, B,
and C be three distinct taxa in $S$.  Without loss of generality,
assume that (AB)C is the $j$th rooted triple on $\sigma$.  From
Lemma~2, $P_{\sigma}[\text{(AB)C}] > P_{\sigma}[\text{(AC)B}] =
P_{\sigma}[\text{(BC)A}]$, where the equality holds by symmetry.  Thus
$P_{\sigma}[\text{(AB)C}] = 1/3 + \delta$ and
$P_{\sigma}[\text{(AC)B}] = 1/3 - \delta/2$ for some $\delta > 0$.  We
use $\widehat{P}$ to denote sample proportions of rooted triples.  For
any $\varepsilon > 0$, because sample proportions converge in
probability to their parametric values (by the Weak Law of Large
Numbers) as the sample size tends to $\infty$, we can choose the
number of loci $k_j$ such that with probability greater than
$1-\varepsilon/\binom{n}{3}$, $\widehat{P}_{\sigma}[\text{(AB)C}] >
1/3$, $\widehat{P}_{\sigma}[\text{(AC)B}] < 1/3$, and
$\widehat{P}_{\sigma}[\text{(BC)A}] < 1/3$.  Letting $k = \max_{j|j
\in \{1,2,\ldots,{n \choose 3}\}} k_j$, for any set of three taxa the
probability that its most common rooted triple in the gene tree
distribution matches the rooted triple in the species tree is
greater than $1-\varepsilon/\binom{n}{3}$.  
The probability that all
of the $\binom{n}{3}$ rooted triples in the $R^*$ tree are
rooted triples in the species tree is therefore greater than
$1-\varepsilon$. $\square$

\begin{center}
\emph{Greedy Consensus}
\end{center}

Three taxa.---For the case of three taxa, greedy consensus applied to
gene trees is asymptotically guaranteed to result in the species tree
as the number of gene trees increases.  If the species tree has
topology ((AB)C) and the one internal branch has length $T$, a random
gene tree has clade (AB) with probability $1-(2/3)e^{-T} > 1/3$,
whereas (AC) and (BC) each occur with probability less than 1/3.  Thus
(AB) is always the most probable cherry for this topology, and the
GACT always matches the species tree topology.  For finitely many
loci, greedy and $R^*$ consensus are not equivalent because they
handle ties differently, with the $R^*$ consensus tree sometimes being
unresolved.

Four taxa.---For the four-taxon 
symmetric species tree and for any choice of branch lengths, 
the GACT has the same topology as the species tree (Appendix~2).
If the species tree is (((AB)C)D), then the GACT
can be the symmetric tree ((AB)(CD)).

To find the set of branch lengths for which the GACT fails to match
the asymmetric species tree topology, let $x$ and $y$ be the lengths of the
deeper and more recent internal branches, respectively, for the tree
(((AB)C)D) (see Fig.~1A).  For this species tree, the region where the GACT is
((AB)(CD)), the ``too-greedy'' zone, consists of those
values of $x$ and $y$ for which the clade $\{\text{CD}\}$ is more
probable than the clade $\{\text{ABC}\}$ (see Appendix~2).  The values
of $x$ and $y$ for which $P(\{\text{CD}\}) > P(\{\text{ABC}\})$
are characterized by
\begin{equation}\label{E:5}
y < \log\left[\frac{3e^{2x}-2}{18(e^{3x}-e^{2x})}\right].
\end{equation}

\noindent The right-hand side of this inequality is strictly less than the
boundary of the anomaly zone for the tree (((AB)C)D) \citep[Equation
(4)]{degnan2006}; thus for this tree, the too-greedy zone is a subset of
the anomaly zone (Fig.~\ref{F:greedyzone}).

More than four taxa.---The result that greedy consensus
can be misleading in the four-taxon case generalizes
to any species topology with more than four taxa.  Intuitively, by
making some branches long and some short (so that coalescent events
occur with probability arbitrarily close to 0 or 1), trees with five
or more taxa can be made to behave similarly to the four-taxon
asymmetric case.  The strategy of the proof is therefore similar to
that of Lemma~5 in Degnan and Rosenberg~(2006).

\vspace{.25in} \textbf{Theorem 5}.  For three-taxon species
topologies, and for four-taxon symmetric species topologies, 
the GACT matches the species tree; for the asymmetric topology with
$n=4$ taxa and for every species topology with $n \geq 5$ taxa, greedy
consensus is inconsistent.
\vspace{.25in}

\vspace{.25in} \textbf{Lemma 6}.  The four-taxon asymmetric topology
(((AB)C)D) has a set of branch lengths which makes greedy consensus fail
to match the species tree.
\vspace{.25in}

\emph{Proof}.  This set is explicitly derived in Appendix~2 and is given in
equation~(5) and Figure~\ref{F:greedyzone}. $\square$



\vspace{.25in}
\textbf{Lemma 7}.  For every bifurcating species tree with $n \geq 5$ taxa 
and every $k \geq 1$ with $2^{k+1}< n$, there is a node
with $c$ terminal descendants, where $2^k < c < 2^{k+1}+1$.
\vspace{.25in}

\emph{Proof}.  For all $k$ satisfying $2^{k+1}+1 \leq n$, the root has
$n \geq 2^{k+1}+1$ terminal descendants.  Let $\mathcal{N}_0$ denote the root node, and let $\mathcal{N}_1$ denote
the internal node immediately descended from the root with the larger
number of terminal descendants (choosing arbitrarily
in case of a tie).  Similarly let $\mathcal{N}_2$ be the internal node
(if it exists) immediately descended from $\mathcal{N}_1$ with the
larger number of terminal descendants.  Continue this process until a
node $\mathcal{N}_m$ ($m \geq 0$) is reached which has at least $2^{k+1}+1$ terminal
descendants, but neither of whose immediate descendant nodes has more than
$2^{k+1}$ terminal descendants. Call $\mathcal{N}_m$ the ``minimal node''.
It follows that at least one of the immediate descendant nodes of the minimal node has more than $2^k$ terminal descendants (since otherwise the minimal node
would have at most $2(2^k) < 2^{k+1}+1$ descendants).  Thus at least one
immediate descendant of the minimal node has $c$ terminal descendants with
$2^k < c < 2^{k+1}+1$.  $\square$

\begin{figure*}[!t]
\begin{center}
\includegraphics[width=.98\textwidth,height=.49\textwidth]{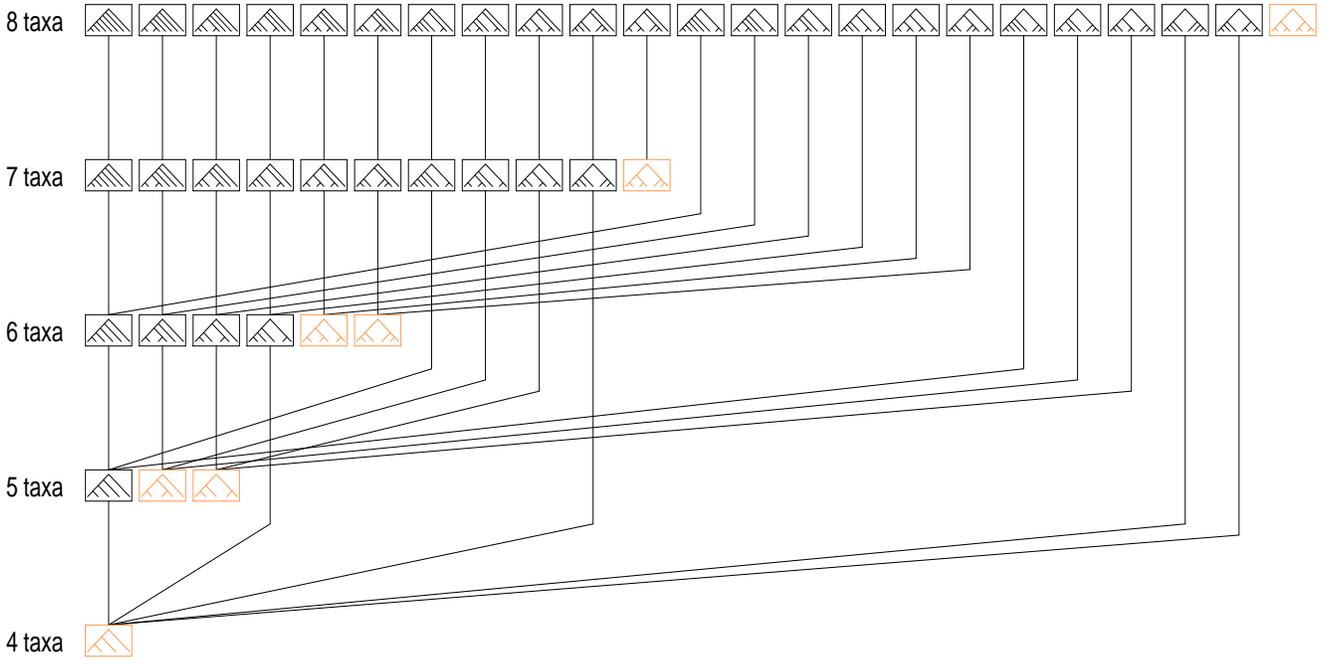}
\end{center}
\caption{Reduction of topologies used in the proof of Lemma 9.  If two
trees are connected by an edge, then the topology with the smaller number
of leaves is a left subtree of the larger tree.}\label{F:figProof}
\end{figure*}

\begin{figure*}
\begin{center}
\includegraphics[width=\textwidth,height=1in]{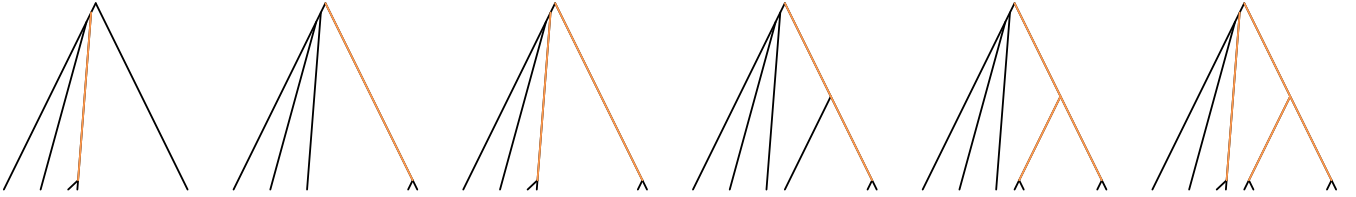}
\end{center}
\caption{Reduction of the remaining trees from Figure 3 to the four-taxon
asymmetric case.  Branches in orange are made long enough that all
lineages on these branches coalesce with probability arbitrarily
close to 1.}\label{F:secondstep}
\end{figure*}


\vspace{.25in}
\textbf{Lemma 8}.  If for some $k \geq 2$, all species tree topologies with $n$ taxa, $n \in \{2^k+1, \dots, 2^{k+1}\}$, have a nonempty too-greedy zone, then all species tree topologies with $n>2{k+1}$ (and thus $n\ge 2^k+1$) taxa have a nonempty too-greedy zone. 

\vspace{.25in}
\emph{Proof}.  Assume there exists $k\ge 2$ such that all species tree
topologies with $n \in \{2^k+1,\dots,2^{k+1}\}$ taxa have a nonempty
too-greedy zone, i.e., that there exist branch lengths for which the
GACT does not match the species tree topology.  By Lemma 7, any
species tree $\sigma$ with more than $2^{k+1}$ ($k \ge 1$) taxa $S$
has some node $\mathcal{N}$ with $c$ terminal descendants, where $c
\in \{2^k+1,\dots,2^{k+1}\}$.  Let $\sigma_{\mathcal{N}}$ denote the
species tree rooted at $\mathcal{N}$ and let $S_{\mathcal{N}}$ denote
the taxa labeling the tips of $\sigma_{\mathcal{N}}$.  By assumption,
$\sigma_{\mathcal{N}}$ has a nonempty too-greedy zone.

Make the lengths of all branches outside of $\sigma_{\mathcal{N}}$
long enough that the probability that all lineages on these long
branches coalesce is greater than $1-\varepsilon$, where $\varepsilon$
is chosen so that $1-\varepsilon > 1/2$ and $1-\varepsilon$ is greater
than the probability of any clade within $\sigma_{\mathcal{N}}$ (i.e.,
any clade which is a proper subset of $S_{\mathcal{N}}$).  Because
the greedy consensus tree is a refinement of the majority-rule consensus
tree,
all clades which include taxa outside of $S_{\mathcal{N}}$, and the
clade consisting of all taxa in $S_{\mathcal{N}}$, are included in the
GACT.  When ranking clade probabilities as is required for the
algorithm for constructing the GACT, these clades are added before the
clades consisting of taxa which are proper subsets of
$S_{\mathcal{N}}$.  Thus eventually the list of candidate clades
consists only of proper subsets of $S_{\mathcal{N}}$.  When clades are
accepted from this list, by assumption we accept at least one
clade to be in the GACT which is not on $\sigma$.  Thus there exist
branch lengths on $\sigma$ for which the GACT does not match the species
tree. $\square$


\vspace{.25in}
\textbf{Lemma 9}.  For any species tree topology with 5, 6, 7, or 8 taxa, there
exists a set of branch lengths for which the greedy asymptotic consensus tree does
not match the species tree.
\vspace{.25in}

\emph{Proof}.  This is shown by reduction to the four-taxon asymmetric
case.  For each species tree topology with 5, 6, 7, or 8 taxa, some
branches can be made long, and some short so as to produce the same
inconsistencies as in the four-taxon case.  Most cases are shown in
Figure~\ref{F:figProof}. Here a topology with $n$ taxa is connected by
an edge to a topology with fewer than $n$ taxa if the smaller topology
is the left subtree---from the node which is the immediate
left-descendant of the root---of the larger topology.  In this case,
for any $\epsilon > 0$, any branches on the larger topology not in the
left subtree can be made arbitrarily long. Thus any lineages
available to coalesce on long branches do coalesce with probability
greater than $1-\varepsilon$.  Remaining clades then have the same
order of probabilities as on the left subtree, and thus are accepted
by the greedy algorithm in the same order as on the left subtree.

If the greedy consensus algorithm returns a nonmatching tree for the
smaller tree, it also does so for the larger tree since the ranking of
the remaining clades by frequencies is eventually the same (once the
high probability clades have already been added on the larger tree).
This process of reducing
trees can be repeated until one of the trees colored orange (which
have no edges connecting to a smaller tree) is reached.

It then remains to be shown that GACT does not match the species tree for
the remaining orange trees from Figure 3.  
This is already shown explicitly for the
four-taxon case (Lemma 6).  For the other trees, these can again
be reduced to the four-taxon case by choosing certain edges to be long
and others short.  This is shown in Figure~\ref{F:secondstep}.  By
choosing the long, orange branches to have large branch lengths, the
probability that all available lineages coalesce on a branch can be
made greater than $1-\varepsilon/(2m)$, where $m$ is the number of long
branches on a tree.  This makes the probability that all available
lineages on long branches coalesce greater than
$1-\varepsilon/2$. Since only counterexamples are needed to show that
the greedy consensus algorithm can return a nonmatching tree, it is
sufficient to note that branches can be chosen to be short enough
using eq.~5 or Figure~5 for the four-taxon asymmetric tree to make the
greedy consensus algorithm fail to return the tree matching the
species tree with probability greater than $1-\varepsilon/2$.  Making
the black internal branches sufficiently short, the probability is
greater than $1-\varepsilon$ that the the entire tree returned by the
greedy consensus algorithm returns fails to match the
species tree topology. $\square$

\vspace{.25in}
\emph{Proof of Theorem 5}.  
The result for three taxa follows from the fact that
the matching gene tree has the highest probability of the three
possible gene trees.  The four-taxon asymmetric case is covered in
Lemma~6. The four-taxon symmetric case is shown to be
consistent in Appendix~2 by showing that for all branch lengths,
(AB) and (CD) are the two most probable clades.  
We have shown that all cases with $n=5,6,7, \text{or } 8$ taxa
have too-greedy zones (Lemma 9).  From Lemma 8, this verifies by induction that all cases with $n \geq 5$ taxa have such zones.
$\square$

\vspace{.25in} \emph{Proof of Theorem 1(ii)}.  The GACT and MACT are
each examples of greedy and majority-rule consensus trees,
respectively.  It follows that if the MACT is fully resolved, then it
is the same as the GACT since greedy consensus trees are refinements
of majority-rule consensus trees \citep{bryant2003}. However, by
Theorem 5, for any species tree topology with $n \ge 5$ taxa, there
exist branch lengths for which the GACT has a clade not on the species
tree, and therefore cannot be equivalent to the MACT (by Theorem
1(i)).  Therefore a sufficient condition for the MACT to be unresolved
is for the GACT to not match the species tree.  Since exact conditions
for the MACT to not be fully resolved were obtained earlier for smaller trees
(the internal branch length being no greater than $\log(4/3)$ for
three-taxon trees and one of eqs.~(1)--(4) to fail for four-taxon
trees), the result follows for any species tree with $n \ge 3$
taxa. $\square$


\begin{figure}[!t]
\begin{center}
\includegraphics[width=.49\textwidth,height=.49\textwidth]{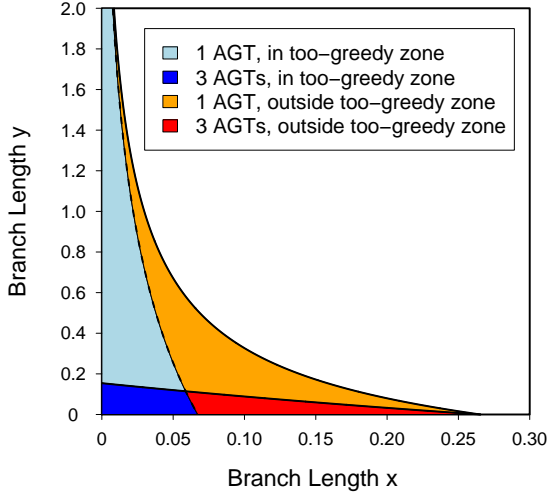}
\end{center}
\caption{The too-greedy zone.  The upper curve is the boundary of
the anomaly zone for the species tree (((AB)C)D).  For points below
this curve, there is either one or three anomalous gene trees (AGTs).  The
two blue regions to the left of the curve which extends from roughly
$(x,y)=(0.067,0.0)$ to $(0.0078,2.0)$ constitute the too-greedy
zone, where the GACT is ((AB)(CD)).}\label{F:greedyzone}
\end{figure}

\begin{center}
{\sc Finite Numbers of Loci}
\end{center}

\begin{center}
\emph{Theory}
\end{center}

The asymptotic consensus trees occur in the limit as the number of
loci approaches infinity.  What happens with a finite number of loci?
In this case, we can examine the behavior of consensus trees from a
theoretical point of view by considering all possible finite samples
of gene trees.  The probability of a particular consensus tree is the
sum of the probabilities of those samples of gene trees that result in
that consensus tree.  These probabilities can be determined by noting
that a sample of independent loci has a multinomial distribution,
where the categories are the gene tree topologies, and the
probabilities are given by the theory of the multispecies coalescent
\citep{degnan2005}.


To compute the probability of a consensus tree given a finite sample
of $\ell$ gene trees, 
let $\ell_i$,$i=1,\dots,k$ be the number of times
gene tree $i$ is observed, where $\sum_i \ell_i = L$, and there are $k$
possible gene tree topologies.
let $c(\ell_1, \dots, \ell_k)$ denote the consensus tree resulting from
a particular sample.  The probability that a sample results in the
consensus tree having topology $\mathcal{T}$ is therefore
\begin{equation}\label{E:cfunction}
\sum_{\substack{\ell_1, \ldots, \ell_k \geq 0\\\ell_1+ \dots +\ell_k=l}}\frac{\ell!}{\ell_1!\cdots
\ell_k!}p_1^{\ell_1} \cdots p_k^{\ell_k}\; I(c(\ell_1, \dots, \ell_k) = \mathcal{T})
\end{equation}
\noindent where $I$ is an indicator that the consensus tree has
topology $T$, $p_i$ is the gene tree probability for the $i$th
topology, and the sum is over all nonnegative integer solutions to
$\ell_1 + \cdots + \ell_k = \ell$.  
There are $\binom{\ell+k-1}{k-1}$
terms in the sum \citep[p.~13]{ross1998}.  For four taxa and
25 loci, the sum has approximately $1.51 \times 10^{10}$ terms.


To compute the probabilities of finite-sample greedy consensus trees,
probabilities of resolutions of ties must also be taken into account.
This can be done by summing over all possible tie-breaks and treating
each possible tie-break as equally likely, rather
than randomly breaking ties.  The
probability of the greedy consensus tree having topology $\mathcal{T}$
can therefore be written as

\begin{equation}\label{E:cfunction2}
\begin{split}
\sum_{\substack{\ell_1, \ldots, \ell_k \geq 0\\\ell_1+ \dots +\ell_k=l}} \frac{\ell!}{\ell_1!\cdots \ell_k!}p_1^{\ell_1} \cdots p_k^{\ell_k} \Biggl[\sum_{b_1 \in B_1} \dots \sum_{b_r \in B_r(b_1,\dots,b_{r-1})} \nonumber\\
 \prod_{j=1}^r \Pr(b_j) I(c(\ell_1, \dots, \ell_k, b_1, \dots, b_r) = \mathcal{T})\Biggr],
\end{split}
\end{equation}

\noindent where $B_j$ denotes the set of possible tie-breaks in the 
$j$th round, $b_j$
denotes one way (out of $|B_j|$ possible ways, where $|B_i|$ is the
number of elements in $B_j$) of breaking up a set of tied clade
frequencies in the $jth$ round (out of $r$ rounds) of choosing clades
for the greedy consensus tree, and $\Pr(b_j) = 1/|B_j|$ is the
probability of a particular tie break. In general, the set $B_j$ is a function
of the choices $b_1, \dots, b_{j-1}$ in preceding rounds of tie-breaks, since
the possible tie breaks in a given round may depend on how previous
ties were resolved.  For $n$-taxon trees, there are $n-2$ rounds of
tie breaks, assuming the case when no tie breaks are necessary (i.e.,
there is one clade on the list which is most frequent) is treated as a
trivial tie break with $|B_j| = 1$.  For example, for four-taxon
trees, there are two rounds of tie breaks.  The function $c$ in
eq.~7 has been given additional arguments (compared with eq.~6) so
that the consensus tree is a function of both the gene tree
frequencies and the tie-breaks.

Because there are a finite number of trees and consensus trees are
computed for every sample, many samples include gene trees which imply
incompatible sets of rooted triples due to there being ties in the
most frequently occurring rooted triple for a given set of taxa.  In
these cases, the $R^*$ algorithm returns a tree which is partially or
completely unresolved.  For example, if there are four input gene
trees: (((AB)C)D), (((AD)C)B), (((BC)A)D), and (((CD)A)B), then the
rooted triples (AD)B and (AB)D each occur twice; thus the $R^*$
consensus tree is unresolved with respect to the relationships between
A, B, and D.  Similarly the rooted triples (BC)D and (CD)B each occur
twice. However, the rooted triple (AC)B occurs twice whereas (AB)C and
(BC)A each occur once, so the $R^*$ consensus tree has the
rooted triple (AC)B.  Similarly, (AC)D occurs in the $R^*$ tree.  Thus
the $R^*$ consensus tree for this set of gene trees is the partially
unresolved tree ((AC)BD).  The majority-rule tree for this
set of input trees is completely unresolved, and the greedy consensus tree
returns each of the four input trees with probability 0.25.

\begin{figure}
\begin{center}
\includegraphics[width=.35\textwidth]{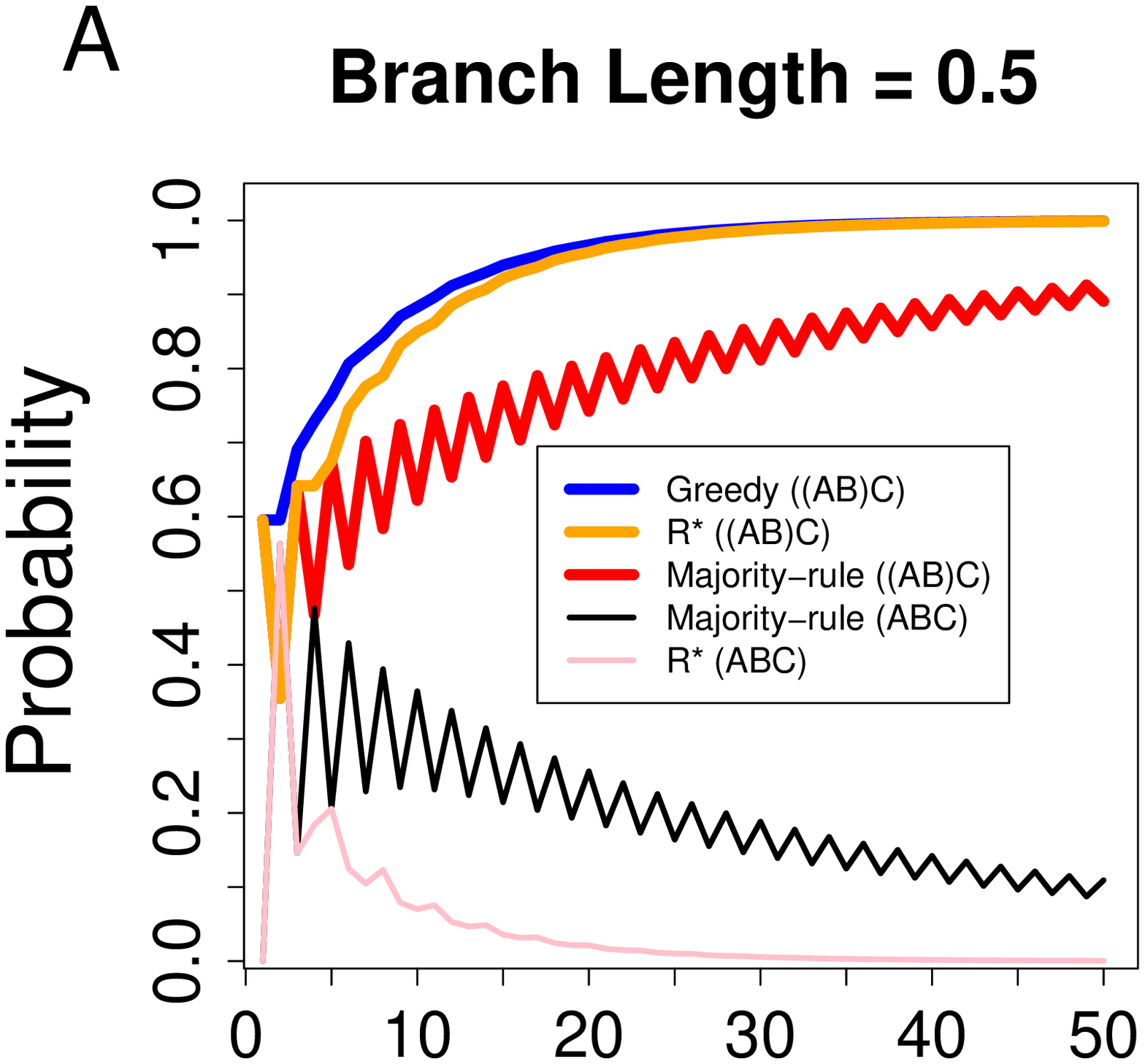}\\
\includegraphics[width=.35\textwidth]{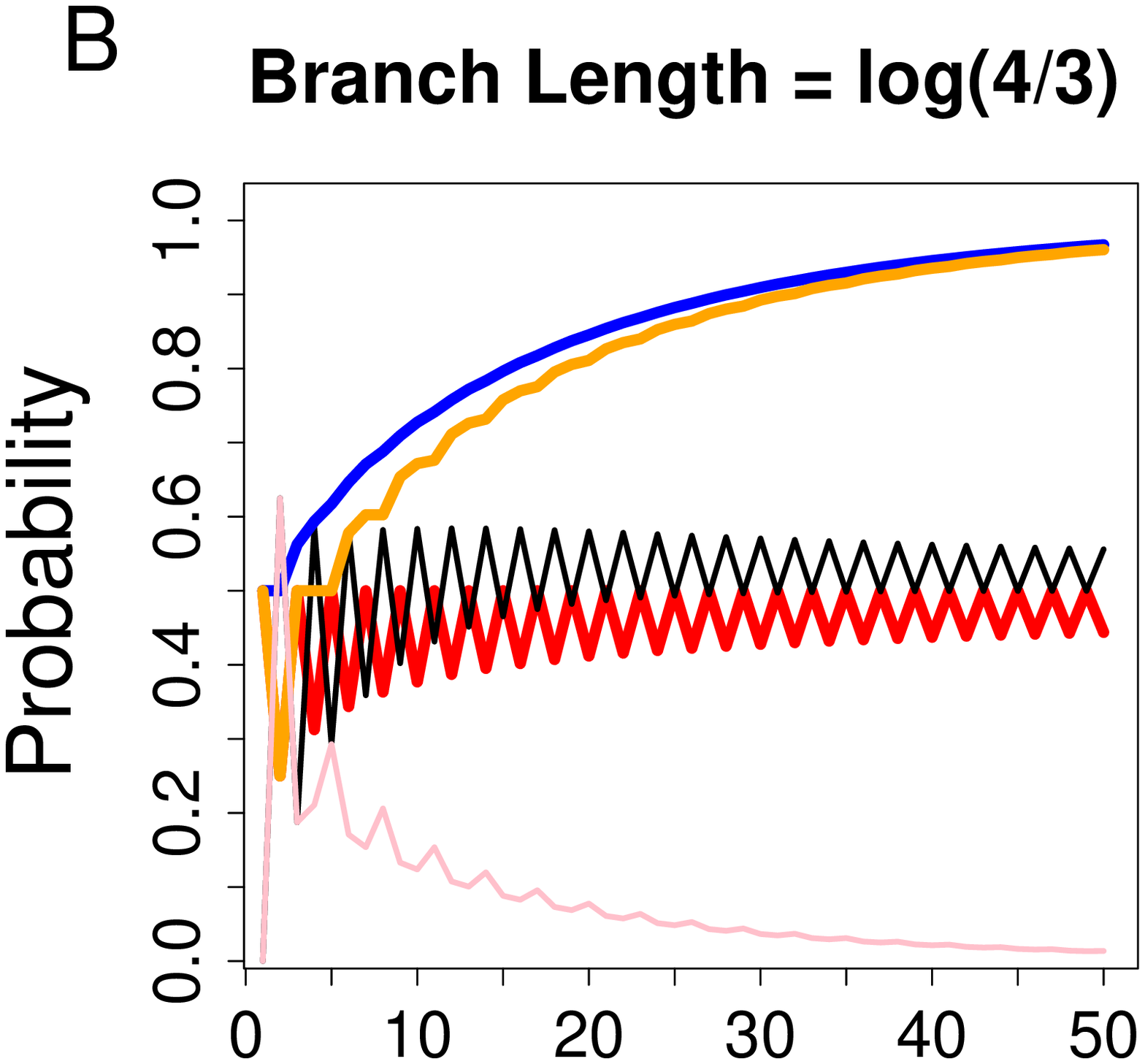}\\
\includegraphics[width=.35\textwidth]{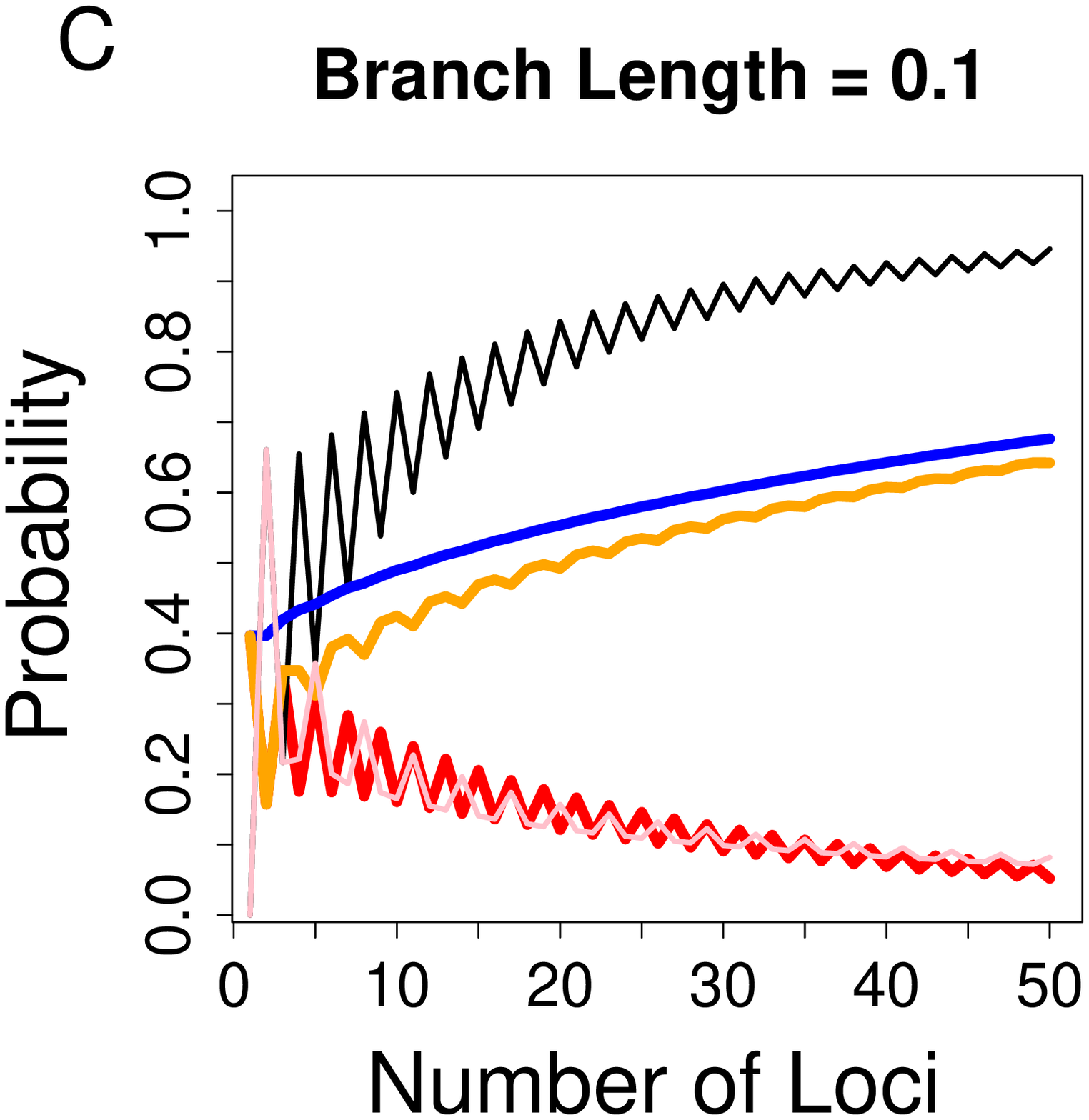}
\end{center}
\caption{Species tree ((AB)C)---Probabilities of consensus trees from
finite numbers of known gene trees.  Each plot shows the probability
that each of the three consensus methods will return either the
species topology, ((AB)C) or a star
tree ($R^*$ and majority-rule only).  
The legend in (A) applies to each of the three plots.}
\label{F:3taxa}
\end{figure}

\begin{center}
\textit{Examples}
\end{center}

\begin{figure*}
\vspace{-1in}
\hspace{-1.5in}
\hspace{.5in}
\includegraphics{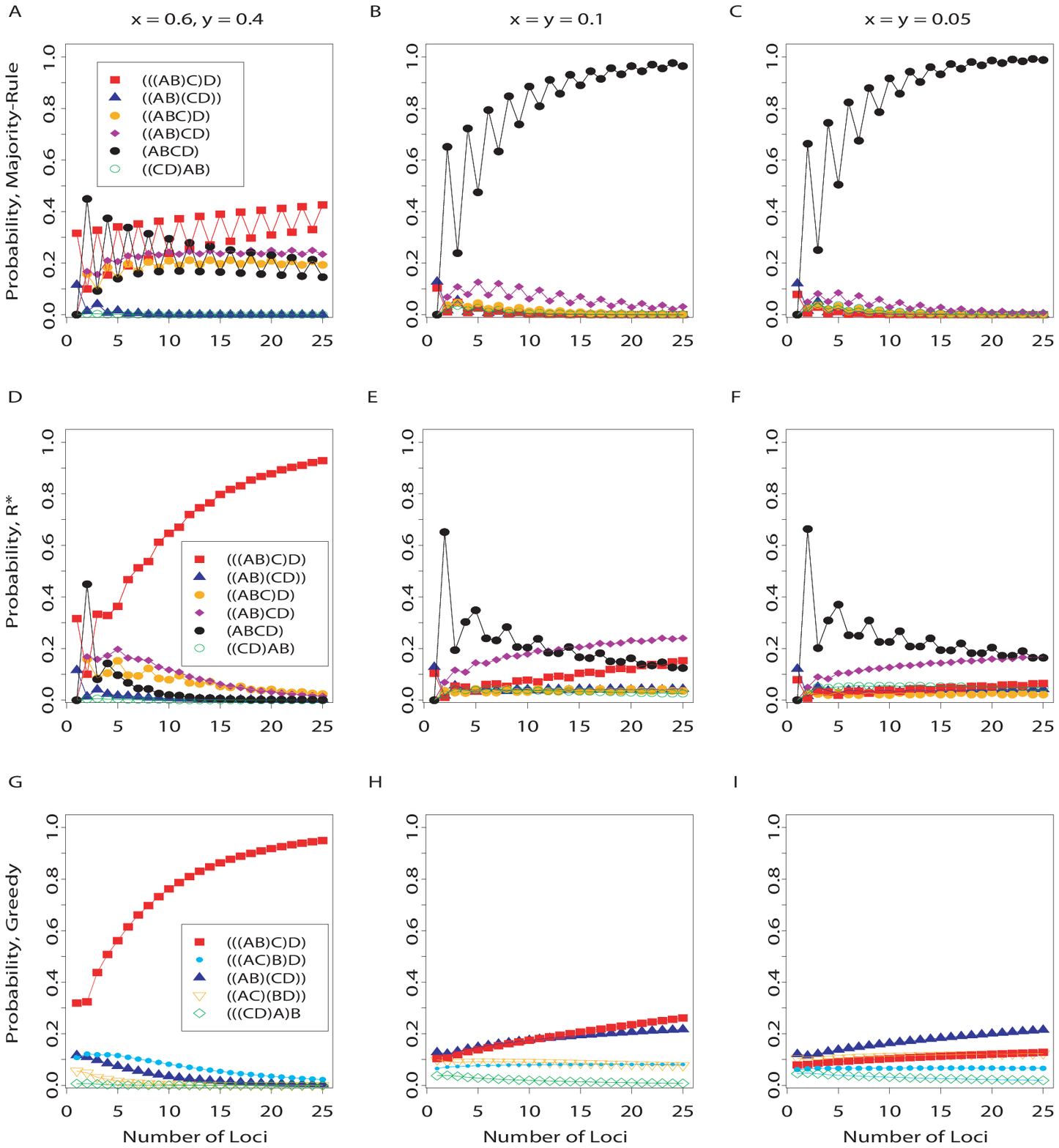}
\vspace{-3in}
\caption{Species tree (((AB)C)D)---Probabilities of consensus trees as
functions of sample size (number of loci).  One consensus algorithm
is used for each row of plots, and one set of branch lengths is used
for each column.  For the majority-rule and $R^*$ algorithms, there
are 26 possible four-taxon consensus trees, including 15 fully resolved
trees and 11 trees not fully resolved.  The graphs only show some of
the more frequently occurring consensus trees; consequently
probabilities do not add to 1.0.  The legends in the lefthand column
apply to the three plots in their corresponding rows.}\label{F:4taxaa}
\end{figure*}

\begin{figure*}
\vspace{-1in}
\hspace{-1.5in}
\hspace{.5in}
\includegraphics{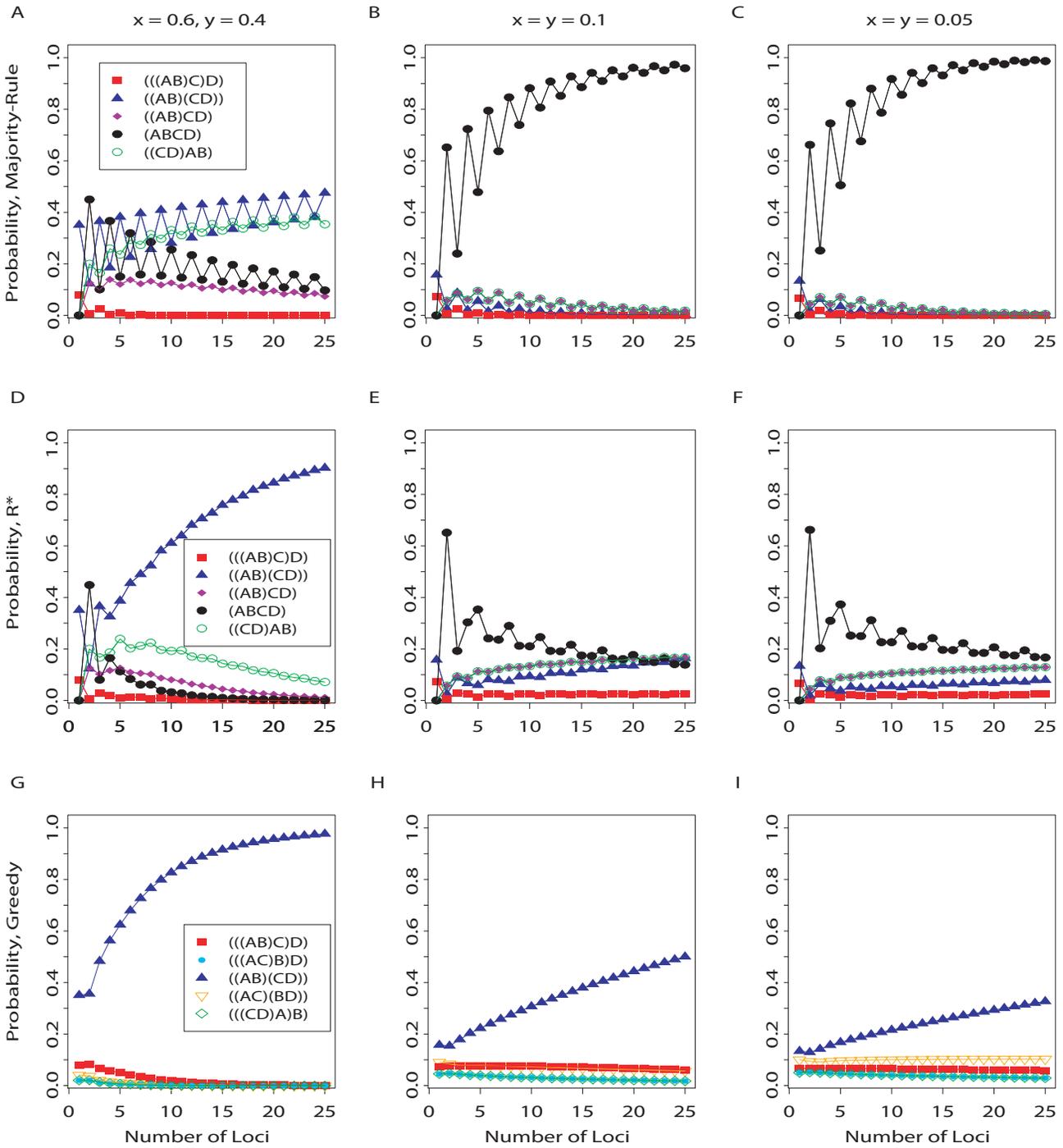}
\vspace{-3in}
\caption{Species tree ((AB)(CD))---Probabilities of consensus trees as
functions of sample size (number of loci).  One consensus algorithm
is used for each row of plots, and one set of branch lengths is used
for each column. For the majority-rule and $R^*$ algorithms, there are
26 possible four-taxon consensus trees, including 15 fully resolved trees
and 11 trees not fully resolved.  The graphs only show some of the
more frequently occurring consensus trees; consequently probabilities
do not add to 1.0.  The legends in the lefthand column apply to the
three plots in their corresponding rows.}\label{F:4taxas}
\end{figure*}

Three taxa.---We illustrate the case of finite loci using three
(Fig.~\ref{F:3taxa}) and four taxa (Figs.~\ref{F:4taxaa} and
\ref{F:4taxas}).  With three taxa, there is only one internal branch
length, and this determines all gene tree probabilities, with the
probability that the gene tree matches the species tree being
$1-(2/3)e^{-T}$, where $T$ is the length of the internal branch. We
used ((AB)C) as the species tree with branch lengths of $0.5,
\log(4/3) \approx 0.288, \text{ and } 0.1$, corresponding to matching
probabilities of 0.596, 0.5, and 0.397, respectively.

For the branch length of 0.5, the majority of loci (almost 60\%)
are likely to have the matching topology; thus, given enough loci, all three methods 
(majority-rule, $R^*$, and greedy) are expected to have
a high probability of returning the matching tree.
This does in fact occur, with the greedy consensus tree having the highest
probability for any given number of loci.  The $R^*$ method has the
second-best performance, although by 50 loci, the greedy and $R^*$ 
algorithms have
roughly equivalent performance.  When the branch length was chosen
such that the probability of matching was 0.5 (Fig.~\ref{F:3taxa}B,
with the two nonmatching trees each having probability 0.25),
majority-rule was stuck between returning the correct tree and the
star tree.  This was not surprising since ((AB)C) by design does not
occur more than 50\% of the time.  The pattern for this
case, as well as for the branch length of 0.1
(Fig.~\ref{F:3taxa}C), continues for greedy and $R^*$ consensus,
with greedy having the best performance, and $R^*$ slowly approaching
greedy as the number of loci increases (and therefore the
probability of ties decreases).  Also, for the branch length of
0.1, no tree has greater than 50\% probability of occurring, and
therefore majority-rule becomes increasingly likely to return a star
tree as the number of loci increases.\\

Four taxa.---Figure~\ref{F:4taxaa} shows the behavior of the
three
consensus methods as the number of loci increases when the species
tree is (((AB)C)D), and Figure~\ref{F:4taxas} shows the same consensus
methods when the species tree is ((AB)(CD)).  The results of the two
figures are similar, although the methods generally perform better with
the symmetric species tree.  

Figure~\ref{F:4taxaa}A suggests that
large numbers of loci might be needed before one majority-rule
consensus tree becomes the most probable.  Figures~\ref{F:4taxaa}B,C and \ref{F:4taxas}B,C, show that majority-rule
consensus can fairly quickly converge to a star phylogeny even though
the probability of a star phylogeny decreases under $R^*$ and greedy
consensus.
     
For majority-rule trees, there is also an effect of having an odd or
even sample size, where even sample sizes tend to give higher
probabilities to unresolved trees.  This occurs because
even sample sizes increase the opportunity for ties in the number
of times two (or more) clades are observed, and in these cases neither
clade can be in the majority.  This has the somewhat surprising result
that a consensus tree can be less likely to match the species tree in
a sample of $2n$ loci than in a sample of $2n-1$ loci (although in
being more likely to have an unresolved tree, it is also less likely
to produce a tree resolved in a way that conflicts with the species
tree).  For the symmetric species topology with branch lengths of
$x=0.6$ and $y=0.4$, note that the majority-rule consensus tree is
more likely to be the species tree topology ((AB)(CD)) than any other
topology if the sample size is odd, but for even sample sizes up to 25
loci, the unresolved tree ((AB)CD) is roughly tied in probability with
((AB)(CD)).  This is consistent with Figure~2B, in which the point
$(x,y)=(0.6,0.4)$ is close the boundary between the regions for
((AB)(CD)) and (AB(CD)).  However, if the number of loci is sufficiently large,
majority-rule consensus is expected to return the resolved
tree ((AB)(CD)) that matches the species topology, since the point
$(x,y) = (0.6,0.4)$ is slightly outside the zone where the MACT is
unresolved.  This can be verified from equations~3 and 4.

As the number of loci increases, the finite-sample $R^*$ trees
(Figs.~\ref{F:4taxaa}D--F and Figs.~\ref{F:4taxas}D--F) show
increasing probability of matching the species tree topology,
regardless of how short the branches are, including for branch lengths
that are in the anomaly zone, $(x,y) = (0.1,0.1)$, and the too-greedy
zone, $(x,y)=(0.05,0.05)$.  This agrees with our theoretical expectations
of $R^*$ consensus trees (Theorem~4); however, the increase in
probability is very slow.  For example, when $(x,y)=(0.1,0.1)$ and the
species tree is asymmetric (Fig.~\ref{F:4taxaa}E), the two trees most
likely to be returned are (ABCD) and ((AB)CD) until there are 23 loci,
at which point the matching topology (((AB)C)D) changes from being the
third to the second most probable topology.  The star tree (ABCD) has
the highest probability for 11 and fewer loci, and as a trend is
decreasing in probability as the sample size increases.  The tree
((AB)CD), however, is still increasing in probability at 25 loci; thus
large numbers of loci might be needed for $R^*$ to show a clear
preference for the matching tree.  The probability that $R^*$ returns
the species tree topology grows more slowly when $(x,y) = (0.05,0.05)$
(Figs.~7F, 8F); however, it is the only one of the three methods 
for which the probability is increasing with those branch
lengths.

Greedy consensus trees 
show more smoothly increasing probabilities of returning the matching
tree for branch lengths outside of the too-greedy zone
(Figs.~\ref{F:4taxaa}G,H and
Figs.~\ref{F:4taxas}G--I).  When the species tree
is (((AB)C)D) and $(x,y)=(0.1,0.1)$ (Fig.~7H), the gene tree ((AB)(CD)) is more
probable than the matching tree, and here greedy consensus is slightly
more likely to return this tree for small samples; but the
matching tree becomes the most probable greedy consensus tree with 11
or more loci.  However, for this species tree, the more extreme branch
lengths of $(x,y)=(0.05,0.05)$ make increasing the number of loci more
likely to result in greedy consensus returning the nonmatching tree ((AB)(CD))
 (Fig.~\ref{F:4taxaa}I).
These results are consistent with our expectations based on the too-greedy zone
(Fig.~\ref{F:greedyzone}).

\begin{center}
{\sc Discussion}
\end{center}

Using coalescent probabilities to determine asymptotic consensus trees 
enables the prediction of what occurs when consensus trees are
constructed from gene trees from many independent loci.  We have
obtained results for the three types of asymptotic consensus trees
considered: majority-rule, $R^*$, and greedy (Theorems 1, 3, and 5,
respectively), which describe the fact that with an infinite number of
loci, MACTs might be unresolved, GACTs might be nonmatching, and RACTs
always match the species tree.  These results have implications for
a common goal of phylogenetics: the inference of species trees.

\begin{center}
\emph{Estimating Species Trees}
\end{center}

Although concatenation of sequences is perhaps the most widely used
method of estimating species trees, there are several current
alternatives to concatenation for inferring species trees.  These
include minimizing deep coalescence \citep{maddison2006}, finding the
joint posterior of the species tree and gene trees from the coalescent
model in a Bayesian framework \citep{liu2007}, using the most ancient
speciation times compatible with the set of inferred coalescent times
on a set of gene trees (called the ``maximum tree'' by Liu and
Pearl~(2007) or ``GLASS tree'' by Mossel and Roch~(2007)), and using
probabilities of gene tree topologies to approximate the species
likelihood \citep{carstens2007,carling2008}.  \nocite{mossel2007}
These methods are designed estimate species trees when there is gene
tree conflict due to incomplete lineage sorting, and they do not
assume that sequence data are generated under a single gene tree
topology.

Theorem~4 suggests a statistically consistent method for building
species tree topologies from gene tree topologies (assuming known gene
trees).  This involves inferring all rooted triples and then applying
a method such as that of Bryant and Berry~(2001) to build up the tree by the
$\binom{n}{3}$ rooted triples.\nocite{bryant2001}

Although this method does not estimate branch lengths on the species
tree, rooted triples could also be used to estimate internal branch
lengths on the species tree by using $P_{\sigma}[\text{(AB)C}] =
1-(2/3)e^{-T}$, where $T$ is the length separating the MRCA of A,
B, and C from the MRCA of A and B.  Thus, the frequency of
each rooted triple in the observed set of gene trees could be used to
estimate species divergence times, from which the species tree
(including topology) could be constructed; or, given a species tree
topology, the set of branch lengths most compatible with the observed
rooted triples could be determined using a criterion such as maximum
likelihood or least squares.

Using majority-rule trees to estimate species trees from finitely many loci 
is expected to not be
misleading, but is likely to result in a tree that is at least partially
unresolved.  It is thus expected to be a conservative estimate of the
species tree, with little power to resolve some clades for some sets of branch lengths.

\phantom{a}

\begin{center}
\emph{Mutation and recombination}
\end{center}

In this paper, we have not considered the roles of mutation and recombination
and the
resulting uncertainty that occurs when gene trees are inferred from
sequence data.  When gene trees are estimated and
the underlying species tree has short branches, some gene trees are
expected to not be fully resolved due to insufficient sequence
divergence.  Due to the inherent stochasticity in sequence evolution,
there will also be some incorrectly inferred gene trees.  For finite
numbers of genes, these factors would tend to increase the probability
that majority-rule consensus trees would have some lack of resolution,
whether or not the true MACT was fully resolved.  If the MACT is a
star tree, we speculate that mutation would cause convergence to a star tree
occur more quickly as the number of loci is increased.  If the MACT
does have some resolved clades, then uncertainty in the gene trees
would be expected to increase the number of loci needed to have a high
probability that the majority-rule tree is correctly resolved.  We
expect similar effects for $R^*$ and greedy consensus trees, but
ultimately, the effects of mutation on constructing consensus trees
could be assessed by simulating sequence data for independent gene
trees evolving in the same species tree.

When recombination occurs within genes, different topologies may exist for
different segments within a gene, further complicating the distribution
of site patterns \citep{wiuf2001}.

\begin{center}
\emph{Conclusions}
\end{center}

Our results show that when there is sufficient gene tree discordance
due to incomplete lineage sorting, majority-rule consensus trees can
have a high probability of being at least partially unresolved, and
the probability of being unresolved can approach 1 as the number of
genes increases indefinitely.  However, the MACT is never resolved
incorrectly; that is, it never has a clade not supported on the
species tree.  We therefore describe the MACT as not misleading;
however, it is not consistent, because statistical consistency implies
that an estimator gets arbitrarily close to a parameter (e.g., a fully
resolved species tree) with probability approaching 1 as the sample
size increases.  


The fact that under the multispecies coalescent, $R^*$ trees are
asymptotically guaranteed to be fully resolved and to match the
species tree topology means that the $R^*$ procedure is not only not
misleading, but is also a statistically consistent estimator
of the species tree topology.  This is remarkable considering that
$R^*$ trees (which are defined for any collection of trees) are based
only minimally on a model of species tree-gene tree relationships.
The only feature of the multispecies coalescent model used in proving the 
consistency of the $R^*$ method is the fact that in this model,
three-taxon relationships that occur in the species tree are also
expected to occur in the gene tree distributions.  Thus, although
$R^*$ consensus trees are consistent without explicitly incorporating
gene tree probabilities into its algorithm for constructing trees, the
$R^*$ consensus tree is not necessarily robust to violations of
assumptions in the coalescent, such as the absence of population
structure along ancient internal edges.

Finally, greedy consensus trees can be increasingly likely (as the
number of gene trees increases) to have a topology that differs from
that of the species tree.  Thus greedy consensus trees can be
positively misleading if used as estimators of species trees.
However, for four taxa, the region of parameter space in which greedy
consensus fails to return the true tree---the too-greedy zone---is
smaller than the anomaly zone; hence greedy consensus offers some
robustness to gene tree discordance that may cause other methods to
fail to recover the species tree.  In addition, the greedy consensus
method outperformed our other methods for branch lengths outside of
the too-greedy zone.  To test these consensus methods in practice will
require examining their performance in the presence of mutation (both
from real and simulated sequence data) that can cause gene trees to be
estimated with uncertainty rather than treated as known.  Although in
our results, $R^*$ consensus outperformed majority-rule consensus, for
$R^*$ and greedy consensus there may be a tradeoff between consistency
and speed of convergence, with greedy consensus being the quicker to
converge yet statistically consistent, and with $R^*$ consensus being 
slow to converge yet statistically consistent.

\section*{Acknowledgments}
We thank C.~An\'e, F.~Matsen, E.~Allman, and an anonymous reviewer for comments.  This work was supported by grants from the National Science Foundation (DEB-0716904), the Burroughs Wellcome Fund, and the Alfred P. Sloan Foundation.
M.D. was supported by training grant T32 GM070449.
D.B. was supported by the NZ Marsden fund.

\begin{footnotesize}
\bibliography{bibfile}
\end{footnotesize}

\clearpage

\begin{table*}
\caption{Probabilities of four-taxon gene tree topologies,
clades, and rooted triples for the species tree (((AB)C)D) with different
branch lengths.  A clade (rooted triple) probability is the sum of probabilities of gene tree topologies which have the clade (rooted triple).  
Branch lengths are as in the model species tree in
Figure~1A.  An asterisk indicates
that a clade has probability greater than 1/2, and would
therefore be represented in the MACT.}\label{T:consensus}
\begin{center}
\fontsize{8}{13}\selectfont
\begin{tabular}{c c c c c c c c c c}
\hline\hline
\multicolumn{6}{c}{}\\[-2ex]
& & & &\multicolumn{6}{c}{Branch lengths $(x,y)$}\\
Gene tree & \multicolumn{2}{c}{Probability} & & $(.6,.4)$ & $(.4,.6)$ & $(.8, .3)$  & $(.3, .3)$ & $(.1, .1)$ & $(.05, .05)$\\  
\multicolumn{5}{l}{}\\[-2.1ex]
\multicolumn{5}{l}{}\\[-4.9ex] \hhline{-~-~------}
\multicolumn{5}{l}{}\\[-2.1ex]
(((AB)C)D)& & $p_1$ & &.316&.319&.321&.212&  .104 & .079\\
(((AB)D)C)& & $p_2$ & &.109&.144&.087&.122&  .091 & .075\\
(((AC)B)D)& & $p_3$ & &.107&.069&.140&.081&  .066 & .061\\
(((AC)D)B)& & $p_4$ & &.049&.043&.048&.058&  .062 & .060\\
(((AD)B)C)& & $p_5$ & &.006&.009&.004&.017&  .037 & .045\\
(((AD)C)B)& & $p_6$ & &.006&.009&.004&.017&  .037 & .045\\
(((BC)A)D)& & $p_7$ & &.107&.069&.140&.081&  .066 & .061\\
(((BC)D)A)& & $p_8$ & &.049&.043&.048&.058&  .062 & .060\\
(((BD)A)C)& & $p_9$ & &.006&.009&.004&.017&  .037 & .045\\
(((BD)C)A)& & $p_{10}$ & &.006&.009&.004&.017&  .037 & .045\\
(((CD)A)B)& & $p_{11}$ & &.006&.009&.004&.017&  .037 & .045\\
(((CD)B)A)& & $p_{12}$ & &.006&.009&.004&.017&  .037 & .045\\
((AB)(CD))& & $p_{13}$ & &.115&.153&.094&.139&  .128 & .121\\
((AC)(BD))& & $p_{14}$ & &.055&.052&.052&.075&  .099 & .105\\
((AD)(BC))& & $p_{15}$ & &.055&.052&.052&.075&  .099 & .105\\
\multicolumn{5}{l}{}\\
Clade & & & & & & &\\[.5ex] 
\multicolumn{5}{l}{}\\[-4.9ex]\hhline{-~~~~~~~~~}
\multicolumn{5}{l}{}\\[-2.1ex]
\{AB\}& & $p_1+p_2+p_{13}$ & & \phantom{*}.541* & \phantom{*}.616* & .499& .473 & .322 & .275\\
\{AC\}& &$p_3+p_4+p_{14}$ & & .211 & .165 & .239& .213 & .227 & .226\\
\{AD\}& &$p_5+p_6+p_{15}$ & & .067 & .071 & .059& .108 & .174 & .196\\
\{BC\}& &$p_7+p_8+p_{15}$   &  & .110 & .104 & .103 & .149  & .227 & .226\\
\{BD\}& &$p_9+p_{10}+p_{14}$ & & .067 & .071 & .059 & .108  & .174 & .196\\
\{CD\}& &$p_{11}+p_{12}+p_{13}$ && .128 & .171 & .098 & .172 & .202 & .212\\
\{ABC\}& &$p_1+p_3+p_7$& &  \phantom{*}.530* & .458 & \phantom{*}.601* &.373& .236 & .201\\
\{ABD\}& &$p_2+p_5+p_9$& &  .121 & .162 & .094 &.155& .165 & .166\\
\{ACD\}& &$p_4+p_6+p_{11}$& & .061 & .061 & .055 &.091 & .136& .151\\
\{BCD\}& &$p_8+p_{10}+p_{12}$ && .061 & .061 & .055 &.091 & .136 & .151\\
\multicolumn{5}{l}{}\\
Rooted triple\\[.5ex]
\multicolumn{5}{l}{}\\[-4.9ex]\hhline{-~~~~~~~~~}
(AB)C & & $p_1+p_2+p_5+p_9+p_{13}$ & &.553 & .634 & .506 & .506 & .397 & .366\\
(AC)B & & $p_3+p_4+p_6+p_{11}+p_{14}$ & &.223 & .183 & .247 & .247 & .302  & .317\\
(BC)A & & $p_7+p_8+p_{10}+p_{12}+p_{15}$ & &.223 & .183 & .247 & .247 & .302 & .317\\
(AB)D & & $p_1+p_2+p_3+p_7+p_{13}$ & &.755 & .755 & .778 & .634 & .454 & .397\\
(AD)B & & $p_4+p_5+p_6+p_{11}+p_{15}$ & &.123 & .123 & .111 & .183 & .273 & .302\\
(BD)A & & $p_8+p_9+p_{10}+p_{12}+p_{14}$ & &.123 & .123 & .111 & .183 & .273 & .302\\
(AC)D & & $p_1+p_3+p_4+p_7+p_{14}$ & &.634 & .553 & .700 & .506 & .397 & .366\\
(AD)C & & $p_2+p_5+p_6+p_9+p_{15}$ & &.183 & .223 & .150 & .247 & .302 & .317\\
(CD)A & & $p_8+p_{10}+p_{11}+p_{12}+p_{13}$ & &.183 & .150 & .247 & .223 & .302& .317\\
(BC)D & & $p_1+p_3+p_7+p_8+p_{15}$ & &.634 & .553 & .700 & .506 & .397 & .366\\
(BD)C & & $p_2+p_5+p_9+p_{10}+p_{14}$ & &.183 & .223& .150 & .247 & .302 & .317\\
(CD)B & & $p_4+p_6+p_{11}+p_{12}+p_{13}$ & &.183 & .223 & .150 & .247 & .302 & .317\\
\hline
\end{tabular}
\end{center}
\end{table*}

\clearpage

\begin{table*}\caption{Probabilities of four-taxon gene trees, clades, and 
rooted triples as functions of
terms $g_{ij}(T)$.  
The branch lengths $x$ and $y$ are as in Figure~1A.
The probabilities of clades (rooted triples) are obtained by adding the probabilities
of gene trees for which have the clade (rooted triple, see
Table~1).  For each entry in the table, the left and right numbers are
the coefficients of the $g_{ij}(T)$ terms for the species trees
(((AB)C)D) and ((AB)(CD)), respectively.}\label{T:consensus3}
\begin{center}
\vspace{-.5cm}
\fontsize{8}{13}\selectfont
\begin{tabular}{c c c c c c c c c }
\hline\hline
\multicolumn{5}{c}{}\\[-2ex]
\multicolumn{5}{l}{}\\[-4.9ex]
\multicolumn{5}{l}{}\\[-2.1ex]
\multicolumn{1}{c}{Gene Tree}    & $g_{21}(y)g_{21}(x)$ & $\frac{1}{3}g_{21}(y)g_{22}(x)$ & $\frac{1}{3}g_{22}(y)g_{21}(x)$ & $\frac{1}{18}g_{22}(y)g_{22}(x)$ & $\frac{1}{3}g_{22}(y)g_{31}(x)$ & $\frac{1}{9}g_{22}(y)g_{32}(x)$ & $\frac{1}{18}g_{22}(y)g_{33}(x)$\\
\multicolumn{5}{l}{}\\[.2ex]
\multicolumn{5}{l}{}\\[-4.9ex] \hhline{--------}
\multicolumn{5}{l}{}\\[-2.1ex]
1. (((AB)C)D) &\hspace{-.1cm} 1,0 &\hspace{-.1cm} 1,1 &\hspace{-.1cm} 0,0 &\hspace{-.1cm} 0,1 &\hspace{-.1cm} 1,0 &\hspace{-.1cm} 1,0 &\hspace{-.1cm} 1,0\\
2. (((AB)D)C) &\hspace{-.1cm} 0,0 &\hspace{-.1cm} 1,1 &\hspace{-.1cm} 0,0 &\hspace{-.1cm} 0,1 &\hspace{-.1cm} 0,0 &\hspace{-.1cm} 1,0 &\hspace{-.1cm} 1,0\\
3. (((AC)B)D) &\hspace{-.1cm} 0,0 &\hspace{-.1cm} 0,0 &\hspace{-.1cm} 0,0 &\hspace{-.1cm} 0,1 &\hspace{-.1cm} 1,0 &\hspace{-.1cm} 1,0 &\hspace{-.1cm} 1,0\\
4. (((AC)D)B) &\hspace{-.1cm} 0,0 &\hspace{-.1cm} 0,0 &\hspace{-.1cm} 0,0 &\hspace{-.1cm} 0,1 &\hspace{-.1cm} 0,0 &\hspace{-.1cm} 1,0 &\hspace{-.1cm} 1,0\\
5. (((AD)B)C) &\hspace{-.1cm} 0,0 &\hspace{-.1cm} 0,0 &\hspace{-.1cm} 0,0 &\hspace{-.1cm} 0,1 &\hspace{-.1cm} 0,0 &\hspace{-.1cm} 0,0 &\hspace{-.1cm} 1,0\\
6. (((AD)C)B) &\hspace{-.1cm} 0,0 &\hspace{-.1cm} 0,0 &\hspace{-.1cm} 0,0 &\hspace{-.1cm} 0,1 &\hspace{-.1cm} 0,0 &\hspace{-.1cm} 0,0 &\hspace{-.1cm} 1,0\\
7. (((BC)A)D) &\hspace{-.1cm} 0,0 &\hspace{-.1cm} 0,0 &\hspace{-.1cm} 0,0 &\hspace{-.1cm} 0,1 &\hspace{-.1cm} 1,0 &\hspace{-.1cm} 1,0 &\hspace{-.1cm} 1,0\\
8. (((BC)D)A) &\hspace{-.1cm} 0,0 &\hspace{-.1cm} 0,0 &\hspace{-.1cm} 0,0 &\hspace{-.1cm} 0,1 &\hspace{-.1cm} 0,0 &\hspace{-.1cm} 1,0 &\hspace{-.1cm} 1,0\\
9. (((BD)A)C) &\hspace{-.1cm} 0,0 &\hspace{-.1cm} 0,0 &\hspace{-.1cm} 0,0 &\hspace{-.1cm} 0,1 &\hspace{-.1cm} 0,0 &\hspace{-.1cm} 0,0 &\hspace{-.1cm} 1,0\\
10. (((BD)C)A) &\hspace{-.1cm} 0,0 &\hspace{-.1cm} 0,0 &\hspace{-.1cm} 0,0 &\hspace{-.1cm} 0,1 &\hspace{-.1cm} 0,0 &\hspace{-.1cm} 0,0 &\hspace{-.1cm} 1,0\\
11. (((CD)A)B) &\hspace{-.1cm} 0,0 &\hspace{-.1cm} 0,0 &\hspace{-.1cm} 0,1 &\hspace{-.1cm} 0,1 &\hspace{-.1cm} 0,0 &\hspace{-.1cm} 0,0 &\hspace{-.1cm} 1,0\\
12. (((CD)B)A) &\hspace{-.1cm} 0,0 &\hspace{-.1cm} 0,0 &\hspace{-.1cm} 0,1 &\hspace{-.1cm} 0,1 &\hspace{-.1cm} 0,0 &\hspace{-.1cm} 0,0 &\hspace{-.1cm} 1,0\\
13. ((AB)(CD)) &\hspace{-.1cm} 0,1 &\hspace{-.1cm} 1,1 &\hspace{-.1cm} 0,1 &\hspace{-.1cm} 0,2 &\hspace{-.1cm} 0,0 &\hspace{-.1cm} 1,0 &\hspace{-.1cm} 2,0\\
14. ((AC)(BD)) &\hspace{-.1cm} 0,0 &\hspace{-.1cm} 0,0 &\hspace{-.1cm} 0,0 &\hspace{-.1cm} 0,2 &\hspace{-.1cm} 0,0 &\hspace{-.1cm} 1,0 &\hspace{-.1cm} 2,0\\
15. ((AD)(BC)) &\hspace{-.1cm} 0,0 &\hspace{-.1cm} 0,0 &\hspace{-.1cm} 0,0 &\hspace{-.1cm} 0,2 &\hspace{-.1cm} 0,0 &\hspace{-.1cm} 1,0 &\hspace{-.1cm} 2,0\\
\multicolumn{5}{l}{}\\
Clade &\hspace{-.1cm} &\hspace{-.1cm} &\hspace{-.1cm} &\hspace{-.1cm} &\hspace{-.1cm} \\
\multicolumn{5}{l}{}\\[.2ex]
\multicolumn{5}{l}{}\\[-4.9ex]\hhline{-~~~~}
\multicolumn{5}{l}{}\\[-2.1ex]
\{AB\}  &\hspace{-.1cm} 1,1 &\hspace{-.1cm} 3,3 &\hspace{-.1cm} 0,1 &\hspace{-.1cm} 0,4 &\hspace{-.1cm} 1,0 &\hspace{-.1cm} 3,0 &\hspace{-.1cm} 4,0\\
\{AC\}  &\hspace{-.1cm} 0,0 &\hspace{-.1cm} 0,0 &\hspace{-.1cm} 0,0 &\hspace{-.1cm} 0,4 &\hspace{-.1cm} 1,0 &\hspace{-.1cm} 3,0 &\hspace{-.1cm} 4,0\\
\{AD\}  &\hspace{-.1cm} 0,0 &\hspace{-.1cm} 0,0 &\hspace{-.1cm} 0,0 &\hspace{-.1cm} 0,4 &\hspace{-.1cm} 0,0 &\hspace{-.1cm} 1,0 &\hspace{-.1cm} 4,0\\
\{BC\}  &\hspace{-.1cm} 0,0 &\hspace{-.1cm} 0,0 &\hspace{-.1cm} 0,0 &\hspace{-.1cm} 0,4 &\hspace{-.1cm} 1,0 &\hspace{-.1cm} 3,0 &\hspace{-.1cm} 4,0\\
\{BD\}  &\hspace{-.1cm} 0,0 &\hspace{-.1cm} 0,0 &\hspace{-.1cm} 0,0 &\hspace{-.1cm} 0,4 &\hspace{-.1cm} 0,0 &\hspace{-.1cm} 1,0 &\hspace{-.1cm} 4,0\\
\{CD\}  &\hspace{-.1cm} 0,1 &\hspace{-.1cm} 1,1 &\hspace{-.1cm} 0,3 &\hspace{-.1cm} 0,4 &\hspace{-.1cm} 0,0 &\hspace{-.1cm} 1,0 &\hspace{-.1cm} 4,0\\
\{ABC\} &\hspace{-.1cm} 1,0 &\hspace{-.1cm} 1,1 &\hspace{-.1cm} 0,0 &\hspace{-.1cm} 0,3 &\hspace{-.1cm} 3,0 &\hspace{-.1cm} 3,0 &\hspace{-.1cm} 3,0 \\
\{ABD\} &\hspace{-.1cm} 0,0 &\hspace{-.1cm} 1,1 &\hspace{-.1cm} 0,0 &\hspace{-.1cm} 0,3 &\hspace{-.1cm} 0,0 &\hspace{-.1cm} 1,0 &\hspace{-.1cm} 3,0\\
\{ACD\} &\hspace{-.1cm} 0,0 &\hspace{-.1cm} 0,0 &\hspace{-.1cm} 0,1 &\hspace{-.1cm} 0,3 &\hspace{-.1cm} 0,0 &\hspace{-.1cm} 1,0 &\hspace{-.1cm} 3,0\\
\{BCD\} &\hspace{-.1cm} 0,0 &\hspace{-.1cm} 0,0 &\hspace{-.1cm} 0,1 &\hspace{-.1cm} 0,3 &\hspace{-.1cm} 0,0 &\hspace{-.1cm} 1,0 &\hspace{-.1cm} 3,0\\
\multicolumn{5}{l}{}\\
Rooted Triple &\hspace{-.1cm} &\hspace{-.1cm} &\hspace{-.1cm} &\hspace{-.1cm} &\hspace{-.1cm}\\
\multicolumn{5}{l}{}\\[.2ex]
\multicolumn{5}{l}{}\\[-4.9ex]\hhline{-~~~~}
\multicolumn{5}{l}{}\\[-2.1ex]
(AB)C &\hspace{-.1cm} 1,1 &\hspace{-.1cm} 3,3 &\hspace{-.1cm} 0,1 &\hspace{-.1cm} 0,6 &\hspace{-.1cm} 1,0 &\hspace{-.1cm} 3,0 &\hspace{-.1cm} 6,0\\
(AC)B &\hspace{-.1cm} 0,0 &\hspace{-.1cm} 0,0 &\hspace{-.1cm} 0,1 &\hspace{-.1cm} 0,6 &\hspace{-.1cm} 1,0 &\hspace{-.1cm} 3,0 &\hspace{-.1cm} 6,0\\
(BC)A &\hspace{-.1cm} 0,0 &\hspace{-.1cm} 0,0 &\hspace{-.1cm} 0,1 &\hspace{-.1cm} 0,6 &\hspace{-.1cm} 1,0 &\hspace{-.1cm} 3,0 &\hspace{-.1cm} 6,0\\
(AB)D &\hspace{-.1cm} 1,1 &\hspace{-.1cm} 3,3 &\hspace{-.1cm} 0,1 &\hspace{-.1cm} 0,6 &\hspace{-.1cm} 3,0 &\hspace{-.1cm} 5,0 &\hspace{-.1cm} 6,0\\
(AD)B &\hspace{-.1cm} 0,0 &\hspace{-.1cm} 0,0 &\hspace{-.1cm} 0,1 &\hspace{-.1cm} 0,6 &\hspace{-.1cm} 0,0 &\hspace{-.1cm} 2,0 &\hspace{-.1cm} 6,0\\
(BD)A &\hspace{-.1cm} 0,0 &\hspace{-.1cm} 0,0 &\hspace{-.1cm} 0,1 &\hspace{-.1cm} 0,6 &\hspace{-.1cm} 0,0 &\hspace{-.1cm} 2,0 &\hspace{-.1cm} 6,0\\
(AC)D &\hspace{-.1cm} 1,0 &\hspace{-.1cm} 1,1 &\hspace{-.1cm} 0,0 &\hspace{-.1cm} 0,6 &\hspace{-.1cm} 3,0 &\hspace{-.1cm} 5,0 &\hspace{-.1cm} 6,0\\
(AD)C &\hspace{-.1cm} 0,0 &\hspace{-.1cm} 1,1 &\hspace{-.1cm} 0,0 &\hspace{-.1cm} 0,6 &\hspace{-.1cm} 0,0 &\hspace{-.1cm} 2,0 &\hspace{-.1cm} 6,0\\
(CD)A &\hspace{-.1cm} 0,1 &\hspace{-.1cm} 1,1 &\hspace{-.1cm} 0,3 &\hspace{-.1cm} 0,6 &\hspace{-.1cm} 0,0 &\hspace{-.1cm} 2,0 &\hspace{-.1cm} 6,0\\
(BC)D &\hspace{-.1cm} 1,0 &\hspace{-.1cm} 1,1 &\hspace{-.1cm} 0,0 &\hspace{-.1cm} 0,6 &\hspace{-.1cm} 3,0 &\hspace{-.1cm} 5,0 &\hspace{-.1cm} 6,0\\
(BD)C &\hspace{-.1cm} 0,0 &\hspace{-.1cm} 1,1 &\hspace{-.1cm} 0,0 &\hspace{-.1cm} 0,6 &\hspace{-.1cm} 0,0 &\hspace{-.1cm} 2,0 &\hspace{-.1cm} 6,0\\
(CD)B &\hspace{-.1cm} 0,1 &\hspace{-.1cm} 1,1 &\hspace{-.1cm} 0,3 &\hspace{-.1cm} 0,6 &\hspace{-.1cm} 0,0 &\hspace{-.1cm} 2,0 &\hspace{-.1cm} 6,0\\
\hline
\end{tabular}
\end{center}
\end{table*}

\clearpage


\clearpage
{\sc Appendix 1: Majority-Rule Unresolved Zones, Species Tree (((AB)C)D)}
\vspace{.25in}

In this appendix we derive conditions for which the MACT is unresolved 
for the four-taxon species trees (((AB)C)D) and ((AB)(CD)).
This is done by finding branch lengths for which there exist clades
with probability greater than 1/2.
First, the following result about cherries 
is useful, which is analogous to Proposition~1 and has a similar
proof.

\vspace{.25in} 
\textbf{Proposition 10.} Let $\sigma$ be the species tree where $S$ is
the set of taxa on $\sigma$.  Then for any A, B, C $\in S$, if
\{AB\} is a cherry on $\sigma$, then $P_{\sigma}[\text{\{AB\}}] >
P_{\sigma}[\text{\{AC\}}]$.
\vspace{.25in}

\emph{Proof}.  The proof is very similar to the proof of Lemma 2.


\vspace{.25in}
\textbf{Remark 11}. If \{AB\} is a cherry on the species tree $\sigma$,
then for any taxon C, $P_{\sigma}[\text{\{AC\}}] = P_{\sigma}[\text{\{BC\}}] < 1/3$.
\vspace{.25in}

\noindent The equality holds by symmetry; the inequality follows from
Proposition~10.

To find branch lengths for the species tree (((AB)C)D) where the MACT is
resolved, consider the probabilities of clades \{ABC\} and \{AB\}.
Table~\ref{T:consensus}
lists the probability that A, B, and C are monophyletic as $p_1 +
p_3 + p_7$, where $p_i$ is the probability of gene tree
$i$ in the same table, because for gene trees 1, 3, and 7 (and only these gene trees),
these three taxa are monophyletic.  Table~2 can be used to compute probabilities of gene trees, clades, or rooted triples for four-taxon trees as linear
combinations of products of the terms $g_{i,j}(T)$, 
which denote the probability
that $i$ lineages coalesce into $j$ lineages within $T$ coalescent units,
where $i \geq j \geq 1$, and $T > 0$.  For $i=2,3$, the $g_{ij}(t)$ functions
are \citep{tavare1984,pamilo1988}:
\begin{alignat}{2}\label{E:gijoe}
g_{21}(T) &= 1-e^{-T} \hspace{.1in} && g_{31}(T) = 1 - \frac{3}{2}e^{-T} + \frac{1}{2}e^{-3T}\notag\\
g_{22}(T) &= e^{-T}  && g_{32}(T) = \frac{3}{2}e^{-T} - \frac{3}{2}e^{-3T}\\
& &&g_{33}(T) = e^{-3T}
\end{alignat}

For example, we see from Table~2 that if the species
tree is (((AB)C)D), the probability of clade $\{\text{CD}\}$ is
$\frac{1}{3}g_{21}(y)g_{22}(x) + \frac{1}{9}g_{22}(y)g_{32}(x) +
\frac{4}{18}g_{22}(y)g_{33}(x)$; and if the species tree is
((AB)(CD)), the probability of clade $\{\text{CD}\}$ is
$g_{21}(y)g_{21}(x) + \frac{1}{3}g_{21}(y)g_{22}(x) +
\frac{1}{3}g_{22}(y)g_{21}(x) + \frac{4}{18}g_{22}(y)g_{33}(x)$.

\begin{align}
P_{\sigma}[\{\text{ABC}\}] &= p_1 + p_3 + p_7 \notag\\ 
&= 1 - \frac{2}{3}e^{-x}-\frac{1}{3}e^{-(x+y)}+\frac{1}{6}e^{-(3x+y)}.
\end{align}
Setting $P_{\sigma}[\{\text{ABC}\}] > 1/2$, we obtain a condition for
which the consensus tree has the clade \{ABC\}.  We also note that
no other three-taxon clade can be on the MACT because they are each
incompatible with and less probable than \{ABC\}, and therefore have
probabilities less than 1/2.  This can be verified by checking their
probabilities from Table~2 and comparing coefficients of the $g_{ij}(T)$ terms,

Three-taxon clades for the species tree (((AB)C)D) have the probabilities:
\begin{align*}
P_{\sigma}(\text{\{ABC\}}) &= g_{21}(y)g_{21}(x) + \frac{1}{3}g_{21}(y)g_{22}(x) \\
&+ g_{22}(y)g_{31}(x) + \frac{3}{9}g_{22}(y)g_{32}(x)\\
&= \frac{3}{18}g_{22}(y)g_{33}(x) \\
P_{\sigma}(\text{\{ABD\}}) &= \frac{1}{3}g_{21}(y)g_{22}(x) + \frac{1}{9}g_{22}(y)g_{32}(x)\\
&+ \frac{3}{18}g_{22}(y)g_{33}(x)\\
P_{\sigma}(\text{\{ACD\}}) &= P_{\sigma}(\text{\{BCD\}}) =  \frac{1}{9}g_{22}(y)g_{32}(x)\\ 
&+ \frac{3}{18}g_{22}(y)g_{33}(x)
\end{align*}

The grouping \{AB\} is monophyletic with probability greater than 1/2 if $p_1 + p_2 + p_{13} \
> 1/2$.  Again using Table~2 and eq.~\ref{E:gijoe}, this occurs when
\begin{equation}
P_{\sigma}[\{\text{AB}\}] = 1 - \frac{2}{3}e^{-y}-\frac{1}{9}e^{-(3x+y)}
\end{equation}
is greater than one-half.  Solving for $y$ yields Equation~(\ref{E:2}).

The four trees shown in Figure~2 are the only consensus trees possible
regardless of the set of branch lengths.  To show that
Proposition~10 guarantees that all cherries
incompatible with \{AB\} (which includes all two-taxon
clades other than \{AB\} and \{CD\}) are
less probable than \{AB\} and therefore have probabilities lower than
1/2 and thus cannot be on the MACT.  To show that \{CD\} cannot occur on
the MACT for this species tree, it must be shown
that this clade has probability less than one-half.

The probability that \{CD\} is monophyletic is
\begin{align}
p_{11}+p_{12}+p_{13} &=
\frac{1}{3}e^{-x}-\frac{1}{6}e^{-(x+y)}+\frac{1}{18}e^{-(3x+y)}\notag\\
&< \frac{1}{3} + \frac{1}{18}e^{-(3x+y)} < \frac{1}{3} +
\frac{1}{18} < \frac{1}{2}.
\end{align}

\vspace{.25in}
{\sc Appendix 2: Majority-Rule Unresolved Zones, Species Tree ((AB)(CD))}
\vspace{.25in}

Similar calculations as in Appendix~1 can be performed when the
species tree is ((AB)(CD)).  For this tree, three-taxon groups cannot
have probability greater than 1/3.  For example, the probability for
monophyly of \{ABC\} is (from Table~2 and eq.~\ref{E:gijoe})
\begin{equation}
\frac{1}{3}e^{-x}-\frac{5}{18}e^{-(x+y)} < \frac{1}{3}e^{-x} < \frac{1}{3}.
\end{equation}
Thus the MACT for a symmetric four-taxon species
tree cannot have a clade with three taxa.

All cherries other than \{AB\} and \{CD\} 
are incompatible with these two cherries (which occur on this species tree),
and from Remark~11, any two-taxon clades other than \{AB\} and \{CD\}
have probability less than 1/2 and cannot occur
on the MACT.  The two clades that can occur on the MACT have probabilities

\begin{align}
P_{\sigma}(\{\text{AB}\}) &= 1 - \frac{2}{3}e^{-y}-\frac{1}{9}e^{-(x+y)}, \text{ and} \\
P_{\sigma}(\{\text{CD}\}) &= 1 - \frac{2}{3}e^{-x}-\frac{1}{9}e^{-(x+y)}.
\end{align}
Setting these functions to be greater than 1/2 yields
Equations~(\ref{E:3}) and (\ref{E:4}).

Here the probability that \{AB\} is a clade cannot greater than 1/2
for $y \leq \log (4/3)$, and the probability of clade \{CD\} cannot be
greater than 1/2 for $x < \log(4/3)$.  These values form asymptotes on
the graph of the unresolved zone for the symmetric species tree
(Fig. \ref{F:LOC}B).

\vspace{.25in}
{\sc Appendix 3: The Too-Greedy Zone, Species Tree (((AB)C)D)}
\vspace{.25in}

In this appendix, we show that when the species tree has topology
(((AB)C)D), finding the branch lengths for the too-greedy zone is equivalent
to determining the set of branch lengths for which \{CD\} is more probable
than \{ABC\}.

For the species tree (((AB)C)D) with any set of branch lengths,
\{ABC\} is the most probable three-taxon clade, and \{AB\} is the most
probable two-taxon clade.  These facts can be verified by comparing
clade probabilities in Table 2. 

In general, \{AB\} is not more
probable than \{ABC\}, however, since the branch ancestral to A and B
but not C might be very short and the branch ancestral to A, B, and C,
but not D, might be very long.  In the latter case \{ABC\} has probability
near 1, and \{AB\} has probability near 1/3.

To show that when the species tree has topology (((AB)C)D), the
GACT is always nonmatching if and only if \{CD\} is more
probable than \{ABC\}, we consider cases where \{ABC\} is either (i) more
probable, (ii--iv) less probable, or (v) equally probable as \{AB\}.
In (ii--iv), we also consider whether \{CD\} is (ii) less probable,
(iii) more probable, or (iv) equally probable as \{ABC\}.  Since these
cases exhaust all possibilities, and greedy consensus returns a
nonmatching tree in case (iii) and with probability
1/2 in case (iv), we get the desired result.

(i) $P[\text{\{ABC\}}] > P[\text{\{AB\}}]$.  Here \{ABC\} is the most
probable clade other than \{ABCD\} and is therefore included in the
GACT.  The remaining compatible clades are \{AB\}, \{AC\} and \{BC\}.
By comparing clade probabilities in Table~2, or by using Proposition~10,
\{AB\} is the most probable clade of these three.  Thus the GACT is (((AB)C)D).

(ii) $P[\text{\{CD\}}] < P[\text{\{ABC\}}] < P[\text{\{AB\}}]$.  In this case, \{AB\} is
the most probable clade (other than \{ABCD\}) and is therefore in the
GACT.  The remaining compatible clades are \{CD\}, \{ABC\}, and
\{ABD\}.  Since $P[\text{\{ABD\}}] < P[\text{\{ABC\}}]$ (Table~2), \{ABD\}
cannot be on the GACT, thus
the GACT is (((AB)C)D).  

(iii) $P[\text{\{ABC\}}] < P[\text{\{CD\}}] < P[\text{\{AB\}}]$ In this
case the GACT is ((AB)(CD)).  Also $P[\text{\{ABC\}}] <
P[\text{\{CD\}}] < P[\text{\{AB\}}]$,
so $P[\text{\{ABC\}}] < P[\text{\{CD\}}]$ is a sufficient condition
for the GACT to be ((AB)(CD)). 

(iv) $P[\text{\{ABC\}}] = P[\text{\{CD\}}] < P[\text{\{AB\}}]$
This equality only holds when
eq.~5 is an equality, which is for points on the boundary of the
too-greedy zone.  In this case the GACT is ((AB)(CD)) or (((AB)C)D), each
with probability 1/2.

(v) Finally, if $P[\text{\{ABC\}}] = P[\text{\{AB\}}]$, then the GACT is
(((AB)C)D) since in this case these are the two most probable clades.

Having considered all cases, $P[\text{\{ABC\}}] < P[\text{\{CD\}}]$ 
is necessary and sufficient for ((AB)(CD)) to be the
GACT with probability 1, and $P[\text{\{ABC\}}] = P[\text{\{CD\}}]$
is necessary and sufficient for ((AB)(CD)) to be the GACT with
probability 1/2.  The probabilities of \{ABC\} and  \{CD\} are given in
eqs.~6 and 8, respectively, in Appendix~1.  Setting $P(\{\text{CD}\})
> P(\{\text{ABC}\})$ and solving for $y$ yields eq.~5.

\vspace{.25in}
{\sc Appendix 4: The Too-Greedy Zone, Species Tree ((AB)(CD))}
\vspace{.25in}

We now show that if the species tree has topology ((AB)(CD)), then the
GACT matches the species tree.  First note that for this species tree,
 \{AB\} and \{CD\}
are always each more probable than any three-taxon clade.
This can be verified by comparing coefficients of the $g_{ij}$ terms
in the clade probabilities from Table~2 and by noting that $g_{ij}(T) > 0$ for
$T>0$:

\begin{align*}
P_{\sigma}(\{\text{AB}\}) &= g_{21}(y)g_{21}(x) + \frac{3}{3}g_{21}(y)g_{22}(x)\\
 &\;+ \frac{1}{3}g_{22}(y)g_{21}(x) + \frac{4}{18}g_{22}(y)g_{22}(x)\\
P_{\sigma}(\{\text{CD}\}) &= g_{21}(y)g_{21}(x) + \frac{1}{3}g_{21}(y)g_{22}(x)\\
 &\;+ \frac{3}{3}g_{22}(y)g_{21}(x) + \frac{4}{18}g_{22}(y)g_{22}(x)\\
P_{\sigma}(\{\text{ABC}\}) &= g_{21}(y)g_{21}(x) + \frac{1}{3}g_{21}(y)g_{22}(x)\\
 &\;+\frac{3}{18}g_{22}(y)g_{22}(x)\\
P_{\sigma}(\{\text{ABD}\}) &= \frac{1}{3}g_{21}(y)g_{22}(x)+\frac{3}{18}g_{22}(y)g_{22}(x)\\
P_{\sigma}(\{\text{ACD}\}) &= P_{\sigma}(\{\text{BCD}\}) = P_{\sigma}(\{\text{ABD}\}) 
\end{align*} 
Also, from Proposition 10, \{AB\} is more probable than any cherry
clade other than \{CD\}, and \{CD\} is more probable than any
two-taxon clade other than \{AB\}.  From this it follows that the
first clade chosen in the greedy algorithm (other than \{ABCD\}) is
either \{AB\} or \{CD\}, since any other clade would be less probable
than one of these two.  If \{AB\} is most probable, the remaining
compatible clades are \{CD\}, \{ABC\}, and \{ABD\}.  However, since
\{CD\} is always more probable than \{ACD\} and \{BCD\}, \{CD\} would be chosen
after \{AB\}.  Similarly, if \{CD\} is chosen first, \{AB\} is more probable
than the remaining clades and so is chosen second.  Thus the GACT is always
((AB)(CD)) for this species tree.

\end{document}